\documentclass[12pt,authoryear]{elsarticle}
\usepackage{algorithm}
\usepackage{mdframed}
\usepackage{float}
\usepackage{tabularx}
\usepackage{xcolor}
\usepackage{amsmath}
\usepackage{changepage}
\usepackage{subcaption}
\usepackage{tcolorbox}
\usepackage[noend]{algpseudocode}
\usepackage{float}
\usepackage{amssymb}
\usepackage{multirow}
\usepackage{xcolor}
\usepackage{booktabs}
\makeatletter
\def\mathcolor#1#{\@mathcolor{#1}}
\def\@mathcolor#1#2#3{%
	\protect\leavevmode
	\begingroup
	\color#1{#2}#3%
	\endgroup
}
\makeatother

\newcommand{\ind}[1]{\mathbb{1}\!\left\{#1\right\}}
\newcommand{\RoAS}{\mathrm{RoAS}}
\newif\ifsweeps
\sweepstrue 
\newif\ifrevmark
\revmarkfalse 
\newcommand{\rev}[1]{\ifrevmark{\color{red}#1}\else{#1}\fi}
\newif\ifdatamark
\datamarkfalse 

\newcommand{\revinput}[1]{{\let\origcentering\centering\renewcommand{\centering}{\origcentering\ifrevmark\color{red}\fi}\input{#1}}}
\newcommand{\datinput}[1]{{\let\origcentering\centering\renewcommand{\centering}{\origcentering\ifdatamark\color{blue}\fi}\input{#1}}}

\usepackage{amssymb}
\journal{}
\usepackage{xcolor}
\usepackage{hyperref}
\hypersetup{
	colorlinks= true,
	citecolor=black,
	linkcolor=black, urlcolor=black
}
\AtBeginDocument{\hypersetup{colorlinks=true, linkcolor=black, citecolor=black, urlcolor=black}}
\usepackage{algorithm}
\usepackage{comment}
\usepackage{xstring} 
\AtBeginDocument{
	\let\oldref\ref
	\renewcommand{\ref}[1]{
		\IfBeginWith{#1}{fig:}%
		{{\color{black}Figure~\oldref{#1}}}%
		\IfBeginWith{#1}{eqn:}%
		{{\color{black}\oldref{#1}}}%
		{\IfBeginWith{#1}{tab:}{{\color{black}Table~\oldref{#1}}}{}}}%
}
\usepackage{graphics}
\usepackage{soul}
\usepackage{threeparttable} 
\usepackage{booktabs} 
\usepackage{arydshln}
\usepackage[top=1in, bottom=1in, left=1.25in, right=1.25in]{geometry}
\begin{document}
\newgeometry{top=1.5in, bottom=1in, left=1.25in, right=1.25in}

\begin{frontmatter}
\title{When Predictive Fit Misleads: Selecting Marketing Mix Models Using Experimental RoAS Benchmarks}
\author[measured]{Hao Chen\corref{cor1}}
\ead{hao.chen@stat.ubc.ca}
\author[measured,emory,scimatica]{Alvin Lim}
\cortext[cor1]{Corresponding author}

\address[measured]{Measured Inc., Austin, TX, USA 78701}
\address[emory]{Goizueta Business School, Emory University, Atlanta, GA, USA 30322}
\address[scimatica]{Scimatica LLC, Orlando, FL, USA 32803}

\begin{abstract} \label{sec:abstract}
Marketing Mix Models (MMMs) are used to estimate channel-level Return on Ad Spend (RoAS) and guide marketing investment, yet competing specifications are usually compared on predictive-fit metrics such as out-of-sample $R^2$, MAPE, or RMSE. We show that good predictive fit does not mean good recovery of RoAS. \rev{In simulations with known RoAS, the best-fitting model recovers RoAS well only when all settings, such as adstock and saturation, are correct and sales noise is low. This is possible in synthetic data, but not in real world examples. In a real world setting, a practitioner does not know what the ``correct'' adstock and other settings are, so $R^2$-based selection can produce misleading RoAS estimates. When precise experimental RoAS estimates are available, selecting the specification closest to them gives better RoAS recovery than selecting on fit. Imprecise experiments, however, should not override clear differences in fit. Two applications with geo-holdout experiments show the same pattern.} We propose an applied workflow: calibrate MMM specifications against experimental RoAS estimates when they exist, and otherwise report how RoAS changes across plausible specifications rather than relying on a single best-fitting model.
\end{abstract}

\end{frontmatter}
\pagestyle{empty}

\newpage
\noindent\textbf{Highlights for Practitioners}
\begin{itemize}
    \item Predictive-fit metrics such as $R^2$, MAPE, and RMSE tell how well a model predicts sales, not whether its channel RoAS is right. Models that fit sales almost equally well can give very different RoAS.
    \item In simulations and in real applications with geo-holdout experiments, the best-fitting model often did not give the most accurate RoAS. It works well only when the model settings, such as adstock and saturation, are correct and sales noise is low, and in practice no one knows the correct settings.
    \item When geo-experiments or other lift tests are available, use their RoAS estimates to choose among model specifications, not only to check the final model. Precise experiments improve the choice; imprecise ones should not override a clear difference in fit.
    \item Without experiments, fit several plausible specifications and report how RoAS changes across them, rather than presenting one model's output as the answer.
\end{itemize}
\restoregeometry 
\pagestyle{plain}
\pagenumbering{arabic}

\section{Introduction}

\noindent Marketing Mix Models (MMMs), also known as media mix models, are widely used to measure the effect of marketing spend and to allocate budgets across channels. Their key output is channel-level Return on Ad Spend (RoAS): the incremental sales generated per dollar spent in a channel. Budgets are set on these numbers, so a wrong RoAS leads directly to a wrong allocation. \\

In practice, MMMs are still judged mostly on how well they predict sales. Experiments such as geo-experiments give credible estimates of incremental effects, but they are costly and cannot be run all the time \citep{vaver2011measuring,kerman2017estimating,barajas2020advertising}. Recent studies compare MMM specifications, software, or methods using $R^2$, RMSE, normalized RMSE, temporal-holdout $R^2$, or forecast error, often alongside other criteria \citep{estevez2025primer,tirumala2026deepcausalmmm,gong2024causalmmm,liu2026aienhanced}. These studies show the risk: a good fit is easily taken as evidence that the incremental effects are right, even though the metrics only reward fit to past sales. Multicollinearity, endogeneity, and sensitivity to the specification make this worse \citep{chan2017challenges,ebbes2021dealing,dew2024your}. Popular MMM tools also rely on predictive or posterior predictive checks \citep{robyn,meridian}. Such checks tell whether a model follows sales; they do not tell whether its RoAS is right. For investment decisions \citep{huang2024marketing}, what matters is incremental RoAS, and marginal RoAS for reallocation \citep{farris2010marketing}. \\

A typical analyst has several reasonable specifications that fit sales about equally well but give different channel RoAS. The real question is which model's RoAS can be trusted for budget decisions, not which model predicts sales best. \\

We study this question with simulations and real data. The simulations vary the saturation form, the number of channels and locations, and whether adstock is included; the real data come from two advertisers with geo-holdout experiments. When the model settings are correct and the sales signal is strong, the best-fitting model usually also gives the best RoAS. When sales are noisy or the carryover is wrong, this often fails, and the model chosen by fit gives clearly worse RoAS. In real data an analyst cannot know which of these cases applies. \\

The paper makes three contributions. First, \ifsweeps using a Monte Carlo study with stress tests on spend collinearity, noise, and adstock shape,\else using a Monte Carlo study,\fi{} it shows when the ranking by fit and the ranking by RoAS accuracy diverge: out-of-sample $R^2$ usually picks the best RoAS model when the settings are correct and noise is low, but only in about half of the datasets at high noise and about a third when the carryover is wrong for every candidate. Second, it proposes using experimental RoAS estimates, when they exist, to choose among plausible specifications, and shows when this beats selection on fit and how precise the experiments need to be. Third, when there are no experiments, it shows how to report the uncertainty in RoAS across specifications, and why consensus summaries, such as the median RoAS across specifications, belong in reporting rather than in model selection. \\

Fit checks remain useful: they show whether a model follows the broad sales pattern. But when the goal is channel-level RoAS, fit alone is not enough to choose a model. Experiments, or other credible causal estimates, provide the external check that fit cannot. \\

Section~\oldref{sec:lit} reviews related work. Section~\oldref{sec:motivating} gives a motivating example in which two candidates with practically identical $R^2$ differ by almost half in RoAS error. Section~\oldref{sec:method} sets out the framework. Section~\oldref{sec:performance} reports the simulations: a Monte Carlo study over four saturation forms, two model sizes, and two adstock conditions, followed by \ifsweeps studies of spend collinearity, noise, carryover shape, and noisy experimental benchmarks.\else a study of noisy experimental benchmarks.\fi{} Section~\oldref{sec:real} applies the framework to two advertisers' data with geo-holdout experiments, Section~\oldref{sec:implications} discusses managerial implications, and Section~\oldref{sec:conclusions} concludes.

\section{Related Literature} \label{sec:lit}

Experiments, including geo-experiments, are widely seen as credible measures of the incremental effect of advertising \citep{ding2018causal,vaver2011measuring,kerman2017estimating}, but they are costly \citep{barajas2020advertising} and cannot support continuous measurement. They are often used instead as external benchmarks for MMM RoAS \citep{rukh2023framework}. Our focus on recovering causal parameters also links to the econometric work on identification in marketing: instrumental variables for endogenous price and promotion effects \citep{rossi2014even}, double/debiased machine learning, which combines Neyman-orthogonal scores with cross-fitting so that flexible estimation of nuisance functions does not bias the causal parameter \citep{chernozhukov2018double}, and synthetic control for regional counterfactuals \citep{abadie2010synthetic}. \cite{ebbes2011sense} show that under endogeneity, holdout prediction can favour biased least-squares estimates over consistent instrumental-variable estimates, so validation must match the parameter of interest. We study the same problem for nonlinear MMMs used to report RoAS. \\

MMM is a specialized area of marketing analytics \citep{iacobucci2019state,contreras2026statistical}. Following early work \citep{hanssens1990market}, \cite{hanssens2001market} and \cite{wang2005comparative} established MMM as a way to decompose sales into marketing contributions, usually with frequentist regression and fixed adstock and saturation transforms \citep{chivukula2024role}; later work added hierarchical Bayesian models with sign constraints \citep{chen2021hierarchical}. Multicollinearity is a persistent problem: \cite{ebbes2021dealing} show that highly correlated marketing variables can give different parameter values with similar predictions, and \cite{dew2024your} show that nonlinear and time-varying media effects are hard to tell apart from observational data, that the alternatives imply different allocations, and that designed experiments can separate them. Practitioners have raised the same concern for carryover: \cite{palu2026adstock} reports that adstock kernels of different shapes fit the same data almost equally well but imply channel multipliers that differ several-fold, and argues for richer, transfer-function models of carryover. Carryover, saturation, and attribution have been studied widely \citep{newell2008mass,liu2014media,gaur2020attribution}, and industry authors have pointed to the gap between academic methods and practice \citep{chan2017challenges}. Comparisons of MMM estimators have focused on prediction accuracy \citep{wigren2019marketing}, while the importance of RoAS has been stressed in digital marketing \citep{katiaj2024increasing} and in campaigns generally \citep{farris2010marketing}. \\

Open-source tools have made MMM widely available. Meta's Robyn \citep{robyn} offers automated hyperparameter search, multi-objective model selection, and budget allocation. Google's Meridian \citep{meridian} is fully Bayesian and builds on earlier Bayesian work on carryover and shape effects and on geo-level hierarchical MMM \citep{wang2017hierarchical,jin2017bayesian,sun2017geo}. Both can use experiments: Robyn adds a lift-calibration error as a third objective in its model selection, and Meridian uses experiment results to set channel RoAS priors. \cite{zhang2024calibration} develop such prior calibration and note that practitioners already choose among MMMs by how close their RoAS is to experimental estimates, a choice they regard as subjective. Our contribution is a systematic evaluation of how often, and by how much, fit-based selection departs from selection against experiments in nonlinear MMMs, including when the experiments are noisy. Posterior predictive checks in Bayesian workflows ask whether a model reproduces observed sales; used alone, they share the limitation of $R^2$. Posterior distributions of RoAS do quantify estimation uncertainty, and a Bayesian version of our criterion would compare posterior RoAS with experimental benchmarks. The literature has expanded the available estimators, transforms, and software, but has paid less attention to how competing specifications should be judged when the target is causal RoAS rather than sales prediction. That is our focus.

\section{Motivating Example: \rev{When Better Predictions Do Not Give Better RoAS Estimates}} \label{sec:motivating}
Consider an MMM for a product sold in $L = 8$ retail locations over $T = 208$ weeks (four years), with spend in $C = 6$ channels. Sales are generated as follows:

\begin{align}
\log(y_{lt}) &= \beta_0 + \sum_{j=2}^{L} \beta_{L,j} \ind{l=j} + \sum_{i=1}^{C} \left(\beta_i + \sum_{j=2}^{L} \beta_{i,j} \ind{l=j}\right) \cdot m_{i,lt} \notag \\
&\quad + \beta_p \log(p_{lt}) + \beta_t t + \beta_d d_{lt} + \beta_u u_t + \beta_c c_t \notag \\
&\quad + \sum_{k=1}^{3} \gamma_k h_{k,t} + \beta_s^{(1)} \sin\left(\frac{2\pi t}{52}\right) + \beta_s^{(2)} \cos\left(\frac{2\pi t}{52}\right) + \epsilon_{lt}
\label{eq:dgp}
\end{align}
where $y_{lt}$ is sales units at location $l$ in week $t$, the superscripts (1) and (2) mark the sine and cosine terms of the annual cycle, and the transformed media variable $m_{i,lt}$ is built in two stages:
\begin{equation}
m_{i,lt} = f_{\text{sat}}\left(g_{\text{adstock}}(x_{i,lt}; \alpha_i); K_i, O_i\right)
\label{eq:media_transform}
\end{equation}

\begin{itemize}
    \item Adstock Transformation (Geometric Decay):
\begin{equation}
g_{\text{adstock}}(x_{i,lt}; \alpha_i) = x_{i,lt} + \alpha_i \cdot g_{\text{adstock}}(x_{i,l(t-1)}; \alpha_i), \quad 0 \leq \alpha_i < 1
\label{eq:adstock}
\end{equation}
\noindent with initialization $g_{\text{adstock}}(x_{i,l,0}; \alpha_i) = 0$ for each location.

\item Saturation Transformation (Hill Function):
\begin{equation}
f_{\text{sat}}(z_{i,lt}; K_i, O_i) = \frac{z_{i,lt}^{O_i}}{K_i^{O_i} + z_{i,lt}^{O_i}}, \quad K_i > 0, \, O_i > 0
\label{eq:saturation}
\end{equation}
where $z_{i,lt} = g_{\text{adstock}}(x_{i,lt}; \alpha_i)$ represents the adstocked media exposure.
\end{itemize}

Because the model is on the log scale, RoAS is computed from counterfactual predictions on the sales scale. Let $\widehat{\eta}^{(i)}_{lt}$ be the fitted linear predictor with all channels active, and $\widehat{\eta}^{(-i)}_{lt}$ the fitted predictor when channel $i$ is set to zero and all other covariates stay at their observed values. We use the noise-free predictions
\[
\widehat y^{(i)}_{lt}
=
\exp\!\left(\widehat{\eta}^{(i)}_{lt}\right),
\qquad
\widehat y^{(-i)}_{lt}
=
\exp\!\left(\widehat{\eta}^{(-i)}_{lt}\right),
\]
for fitted models and, with the generating parameters, for the known RoAS. Under homoskedastic Gaussian log errors, the conditional mean multiplies both by the same factor $\exp(\sigma^2/2)$.
Channel-level RoAS is then
\begin{equation}
\widehat{\RoAS}_i
=
\frac{\sum_{l=1}^{L} \sum_{t=1}^{T} \left(\widehat y^{(i)}_{lt} - \widehat y^{(-i)}_{lt}\right) p_{lt}}{\sum_{l=1}^{L} \sum_{t=1}^{T} x_{i,lt}},
\label{eq:RoAS}
\end{equation}
This is an average incremental return over the observed spend path and evaluation window, not the return to an additional dollar. Average RoAS is informative for budgeting, but marginal RoAS, or finite-difference returns around the current budget, bear more directly on reallocation. The sums in Equation~\oldref{eq:RoAS} run over an evaluation window $\mathcal W$, and the counterfactual spend path equals the factual one except that channel $i$ spend is zero in $\mathcal W$. Spend before $\mathcal W$ is kept, so the adstock entering $\mathcal W$ is factual; the pre-sample adstock is zero and the recursion runs through the whole sample without restarting. In the simulations $\mathcal W$ is the 52 test weeks, for both the known and the estimated RoAS, and returns after the last week are not counted; in Section~\oldref{sec:real}, $\mathcal W$ is the test weeks of each experiment. The replication code checks this counterfactual against a hand-calculated case. \\

We split the data into training weeks 1--156 ($N_{\text{train}} = 1{,}248$ observations) and test weeks 157--208 ($N_{\text{test}} = 416$). Four specifications are fitted to the training data. All include geometric adstock (Equation~\oldref{eq:adstock}) and differ only in the saturation curve: Hill or Weibull, each in a C-shape (concave) or S-shape (sigmoidal) form (Section~\oldref{sec:saturation}). The data are generated with a C-shape Hill curve, so the Hill C candidate is correctly specified. On the test weeks we score each candidate on sales prediction (out-of-sample $R^2$, MAPE, RMSE) and on RoAS recovery (MAE and MAPE against the known RoAS).

The dataset is one replication from the Monte Carlo study in Section~\oldref{sec:replicated}, chosen by a rule fixed in advance. Among the 160 S2 datasets with adstock-included candidates, the highest-$R^2$ candidate is not the RoAS-best one in 32; we rank these by the ratio of their RoAS-MAEs and show the one with the (lower) median ratio. It is a typical case of disagreement, not an extreme one; Section~\oldref{sec:replicated} reports how often disagreement occurs. Table~\oldref{tab:true_params} (\oldref{app:motivating}) lists its media parameters and Table~\oldref{tab:variables} (\oldref{app:motivating}) all variables. The known RoAS is computed from the generating parameters with the same counterfactual as the estimates. \\

\begin{figure}[htbp]
    \centering
    \includegraphics[width=\textwidth]{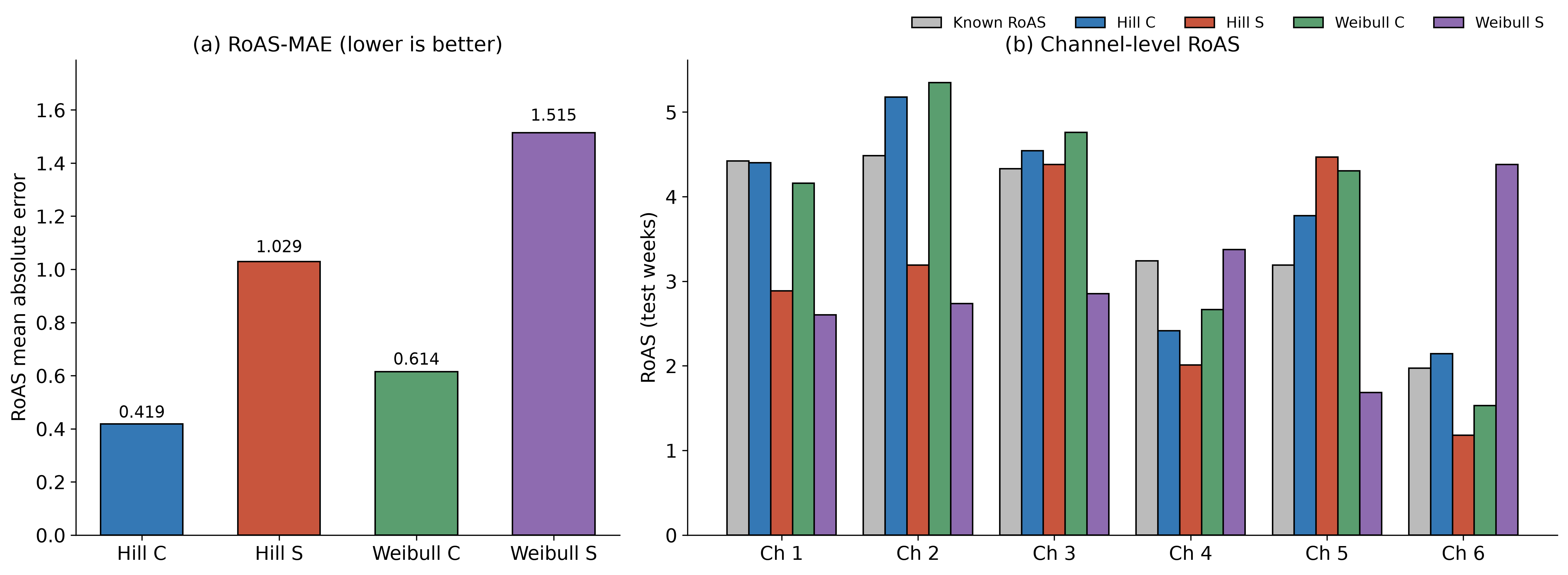}
    \caption{RoAS Recovery of the Four Candidates in the Example Dataset (test weeks). (a) RoAS-MAE against the known RoAS. (b) Channel RoAS estimates and known values. The Weibull C candidate has the highest out-of-sample $R^2$ but 1.47 times the RoAS-MAE of the correctly specified Hill C candidate.}
    \label{fig:RoAS_recovery}
\end{figure}

Table~\oldref{tab:sales_vs_RoAS_accuracy} and Figure~\oldref{fig:RoAS_recovery} (with Figure~\oldref{fig:sales_metrics} in \oldref{app:motivating}) show the problem. The Weibull C model has the highest out-of-sample $R^2$, 0.99480, against 0.99477 for the correct Hill C model. The difference is in the fifth decimal. The two models are also equally close on MAPE (1.70\% against 1.71\%) and RMSE (20.2 against 20.2). Yet the Weibull C model recovers RoAS markedly worse: its RoAS-MAE is 0.614 against 0.419, a ratio of 1.47, and its RoAS-MAPE is 18.4\% against 12.2\%. Both models fit sales equally well but give different RoAS, and nothing in the fit tells which one to trust. \\

The correct model also errs (RoAS-MAPE of 12.2\%), as expected in a finite sample where the adstock and saturation parameters, channel coefficients, location effects, and controls are estimated together. What matters for model choice is the comparison, and there the fit points the wrong way: in a real application the analyst would not know that Hill C is the correct model, and $R^2$ would select Weibull C. The rest of the paper formalizes this evaluation and tests it in simulations and on two advertisers' data with geo-holdout experiments.

{%
\begin{table}[htbp]
\centering
\caption{Sales Prediction and RoAS Recovery of the Four Candidates in the Example Dataset (Test Weeks 157--208)}
\label{tab:sales_vs_RoAS_accuracy}
\small
\begin{tabular}{lcccc}
\toprule
Metric & Hill C & Hill S & Weibull C & Weibull S \\
\midrule
\multicolumn{5}{l}{\textit{Sales prediction}} \\
Out-of-sample $R^2$ & 0.995 & 0.994 & \textbf{0.995} & 0.991 \\
MAPE (\%) & 1.71 & 1.97 & \textbf{1.70} & 2.29 \\
RMSE & 20.2 & 22.5 & \textbf{20.2} & 26.9 \\
\midrule
\multicolumn{5}{l}{\textit{RoAS recovery}} \\
RoAS-MAE & \textbf{0.419} & 1.029 & 0.614 & 1.515 \\
RoAS-MAPE (\%) & \textbf{12.2} & 30.5 & 18.4 & 47.9 \\
\midrule
\multicolumn{5}{l}{\textit{Channel RoAS (known value in parentheses)}} \\
Channel 1 (4.42) & 4.40 & 2.89 & 4.16 & 2.60 \\
Channel 2 (4.48) & 5.18 & 3.19 & 5.35 & 2.74 \\
Channel 3 (4.33) & 4.54 & 4.38 & 4.76 & 2.85 \\
Channel 4 (3.24) & 2.41 & 2.01 & 2.66 & 3.37 \\
Channel 5 (3.19) & 3.78 & 4.47 & 4.31 & 1.69 \\
Channel 6 (1.97) & 2.14 & 1.18 & 1.53 & 4.38 \\
\bottomrule
\end{tabular}
\begin{tablenotes}\footnotesize
\item \textit{Notes}: Bold marks the best value in each row. The data are generated with Hill C saturation; all four candidates include geometric adstock. RoAS is computed over the test weeks (Equation~\oldref{eq:RoAS}).
\end{tablenotes}
\end{table}
}

\section{Experimental-Benchmark Calibration Framework}
\label{sec:method}

An MMM used to report RoAS should be judged against the causal quantity it estimates. The framework below calibrates model choice against experimental or other credible causal RoAS estimates. We set out the model and transformations, then estimation, then RoAS and its evaluation.

\subsection{Model Specification}

We use a multiplicative sales model with location effects. For location $l \in \{1, \ldots, L\}$ and week $t \in \{1, \ldots, T\}$,

\begin{equation}
\label{eq:main_model}
\log(y_{lt}) = \beta_0 + \sum_{j=2}^{L} \beta_{L,j} \ind{l=j} + \sum_{i=1}^{C} \left(\beta_i + \sum_{j=2}^{L} \beta_{i,j} \ind{l=j}\right) \cdot m_{i,lt} + \boldsymbol{X}_{lt}'\boldsymbol{\gamma} + \epsilon_{lt}
\end{equation}

where $y_{lt}$ is sales units, $C$ the number of channels, $m_{i,lt}$ the transformed media variable for channel $i$, and $\boldsymbol{X}_{lt}$ the controls: log price, discounts, macroeconomic indicators, holidays, and seasonality. The error is $\epsilon_{lt} \sim \mathcal{N}(0, \sigma^2)$. The media variable combines carryover (adstock) and diminishing returns (saturation):

\begin{equation}
\label{eq:media_transform2}
m_{i,lt} = f_{\text{sat}}(g_{\text{adstock}}(x_{i,lt};\, \alpha_i);\, K_i, O_i)
\end{equation}

where $x_{i,lt}$ is spend in channel $i$ at location $l$ in week $t$.

\subsection{Transformation Functions}

\subsubsection{Adstock Transformation}

Carryover is modelled by geometric adstock:

\begin{equation}
\label{eq:adstock1}
g_{\text{adstock}}(x_{i,lt}; \alpha_i) = x_{i,lt} + \alpha_i \cdot g_{\text{adstock}}(x_{i,l(t-1)}; \alpha_i), \quad 0 \leq \alpha_i < 1
\end{equation}

with $g_{\text{adstock}}(x_{i,l,0}; \alpha_i) = 0$ for each channel. A larger $\alpha_i$ means a longer-lasting effect.

\subsubsection{Saturation Transformation}\label{sec:saturation}
\begin{figure}[htbp]
    \centering
    \includegraphics[width=0.78\textwidth]{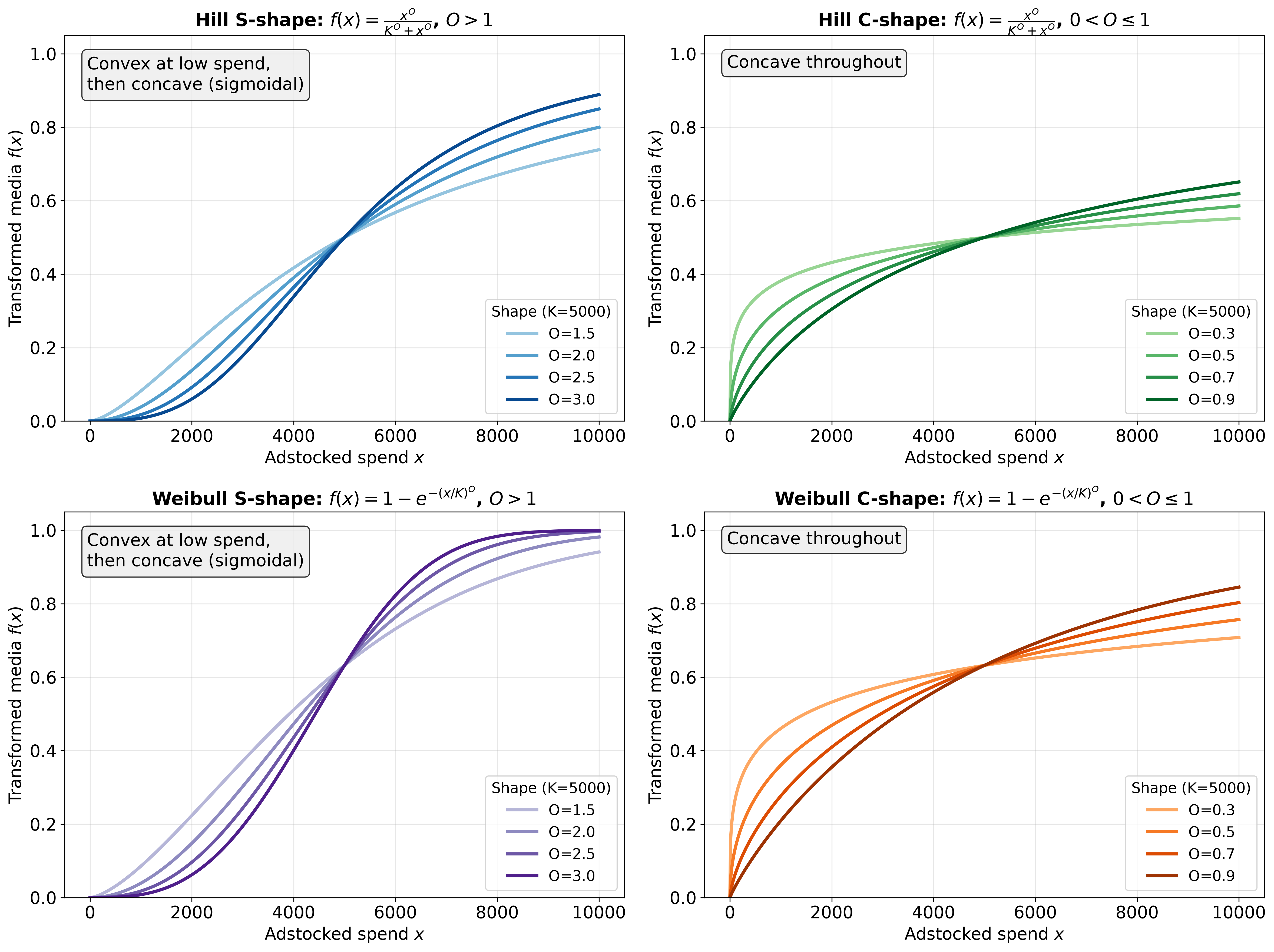}
    \caption{The four candidate saturation curves at scale $K=5000$ and varying shape $O$. Top: Hill S-shape ($O>1$) and Hill C-shape ($0<O\leq 1$). Bottom: Weibull S-shape and Weibull C-shape. S-shapes are convex at low spend and then saturate; C-shapes have high initial returns that fall steadily.}
    \label{fig:saturation_functions}
\end{figure}

Each adstocked exposure $z_{i,lt}=g_{\text{adstock}}(x_{i,lt};\alpha_i)$ is passed through one of two saturation families,
\begin{equation}
\label{eq:sat_families}
f_{\text{Hill}}(z; K, O) = \frac{z^O}{K^O + z^O},
\qquad
f_{\text{Weibull}}(z; K, O) = 1 - \exp\!\left\{-\left(\frac{z}{K}\right)^O\right\},
\end{equation}
with scale $K>0$ and shape $O>0$. For $0<O\leq 1$ both curves are concave everywhere (\emph{C-shape}); for $O>1$ both are sigmoidal (\emph{S-shape}), convex at low spend and then saturating. This gives four candidates (Hill C, Hill S, Weibull C, Weibull S). All are economically plausible, but they imply different average and marginal returns, especially at low spend. These curvature statements refer to $f$, that is, to the log-sales scale. Under the exponential link, expected sales $A\exp\{b f(z)\}$ with $b>0$ are locally convex wherever $b f'(z)^2 > -f''(z)$, even if $f$ is concave, so diminishing returns in sales need further restrictions on the coefficient and the spend range. Figure~\oldref{fig:saturation_functions} shows the four forms; \oldref{app:method} gives their inflection points and compares the two families.

\subsection{Parameter Estimation with Sign Constraints}
Write $\eta_{lt}(\Theta)$ for the right-hand side of Equation~\oldref{eq:main_model} without the error, and assume $\log y_{lt}=\eta_{lt}(\Theta)+\varepsilon_{lt}$ with $\varepsilon_{lt}\stackrel{\mathrm{iid}}{\sim}\mathcal{N}(0,\sigma^2)$. We estimate $\Theta$ by bound-constrained maximum likelihood under sign and shape restrictions: nonnegative channel effects ($\beta_i\geq 0$ and $\beta_i+\beta_{i,j}\geq 0$), $0\leq\alpha_i<1$, $K_i>0$, $O_i\leq 1$ for C-shape and $O_i>1$ for S-shape specifications, and $\sigma>0$. The recursive adstock and nonlinear saturation make the problem non-convex, so we use multi-start L-BFGS-B and keep the solution with the highest likelihood. The coefficients that enter linearly are profiled out exactly, so the numerical search covers only the adstock, saturation, and noise parameters. \oldref{app:method} gives the full likelihood and constraints.

\subsection{Evaluation Framework}

There are two settings, depending on whether experimental RoAS benchmarks exist. The first allows calibration; the second calls for sensitivity reporting.

\subsubsection{Scenario 1: Experimental RoAS Benchmark Available (Primary Framework)}  \label{sec:framework}

When geo-experiments or similar causal studies have been run, their RoAS estimates can serve as calibration targets. The data are split into training (the first $T_{\text{train}}$ weeks) and test ($T_{\text{test}} = T - T_{\text{train}}$ weeks) sets. Each candidate specification (here Hill C, Hill S, Weibull C, and Weibull S) is estimated on the training data by solving Equation~\oldref{eq:ml_constrained}; its RoAS for each channel $i$ is computed on the test set by Equation~\oldref{eq:RoAS}; and the specification with the smallest RoAS recovery error is selected, where
\begin{equation}
\label{eq:RoAS_error}
\text{RoAS-MAE} = \frac{1}{C} \sum_{i=1}^{C} \left| \text{RoAS}_i^{\text{true}} - \text{RoAS}_i^{\text{est}} \right|,
\end{equation}
$\text{RoAS}_i^{\text{true}}$ is the known RoAS in simulations or the experimental estimate in applications, and $\text{RoAS}_i^{\text{est}}$ is the model's estimate.
Algorithm~\oldref{alg:RoAS_framework} summarizes the procedure.

\begin{algorithm}[htbp]
\caption{RoAS-Focused Model Evaluation Framework}
\label{alg:RoAS_framework}
\begin{algorithmic}[1]
\Require Training data $\mathcal{D}_{\text{train}}$, test data $\mathcal{D}_{\text{test}}$, benchmark values $\{\RoAS_i^{\text{true}}\}_{i=1}^{C}$ (known simulation values or experimental estimates), candidate model set $\mathcal{M}$
\Ensure Selected model $m^\star$
\State $\text{BestMAE} \leftarrow \infty$; \ $m^\star \leftarrow \emptyset$
\For{each model $m \in \mathcal{M}$}
    \State Estimate $\hat{\Theta}_m$ on $\mathcal{D}_{\text{train}}$ by solving Equation~\oldref{eq:ml_constrained}
    \State Compute $\widehat{\RoAS}_{i,m}$ on $\mathcal{D}_{\text{test}}$ for $i=1,\dots,C$ (Equation~\oldref{eq:RoAS})
    \State $\mathrm{RoAS\mbox{-}MAE}_m \leftarrow \frac{1}{C}\sum_{i=1}^{C}\bigl|\RoAS_i^{\text{true}} - \widehat{\RoAS}_{i,m}\bigr|$
    \If{$\mathrm{RoAS\mbox{-}MAE}_m < \text{BestMAE}$}
        \State $\text{BestMAE} \leftarrow \mathrm{RoAS\mbox{-}MAE}_m$; \ $m^\star \leftarrow m$
    \EndIf
\EndFor
\State \Return $m^\star$
\end{algorithmic}
\end{algorithm}

\subsubsection{Scenario 2: No Experimental RoAS Benchmark Available}

Experiments are costly and not always feasible. Without a benchmark, fit metrics cannot identify the RoAS-best specification. The right output is then the spread of RoAS across plausible specifications, reported as a sensitivity diagnostic. Section~\oldref{sec:scenario2} examines a simple consensus summary, closeness to the componentwise median RoAS vector, and shows why it should be used for reporting rather than for selection.

\subsection{Comparison to Traditional Evaluation}
MMMs are usually evaluated on sales prediction, using out-of-sample $R^2 = 1 - \sum_{l,t}(y_{lt} - \hat{y}_{lt})^2 / \sum_{l,t}(y_{lt} - \bar{y})^2$, $\text{MAPE} = \frac{100\%}{N}\sum_{l,t} |(y_{lt} - \hat{y}_{lt})/y_{lt}|$, and $\text{RMSE} = \sqrt{\frac{1}{N}\sum_{l,t}(y_{lt} - \hat{y}_{lt})^2}$, where $N$ is the number of test observations. As Section~\oldref{sec:motivating} shows, a model with better sales prediction can give less accurate RoAS. These metrics measure overall fit to observed sales, whereas RoAS also depends on how the adstock and saturation curves extrapolate to the counterfactual spend path. On a fixed sample, $R^2$ and RMSE give the same ranking; MAPE can differ. In the simulations of Section~\oldref{sec:performance}, the rankings diverge most often when the candidates fit almost equally well, when sales are noisy, or when all candidates share a misspecified carryover. The framework therefore keeps two questions apart: fit metrics ask whether a model captures sales patterns, and RoAS diagnostics ask whether it recovers the incremental effects used for decisions. With experiments, the second question can be answered directly; without them, the spread of RoAS across plausible specifications should be reported, not hidden behind one fit-selected model.

In practice the workflow has six steps: fit substantively plausible specifications; compare their fit; if experiments exist, compare each model's RoAS for the same contrast with the experimental estimates (RoAS-MAE, or a discrepancy with precision weights fixed across candidates); if not, report the median, interquartile range, and range of RoAS across specifications; state which channel conclusions are stable and which depend on the specification; and use the calibrated or sensitivity-qualified RoAS for decisions. We use the equal-weight absolute loss in Equation~\oldref{eq:RoAS_error} for transparency; precision-weighted or business-value losses fit into the same framework.

\section{Performance Evaluation}
\label{sec:performance}

We compare ranking by fit with ranking by RoAS accuracy over hundreds of independent simulated datasets. Section~\oldref{sec:sim_design} describes the design. Section~\oldref{sec:replicated} reports a Monte Carlo study\ifsweeps{} and three stress tests (spend collinearity, noise, and carryover shape)\fi{} and evaluates selection against \emph{noisy} experimental benchmarks.

\subsection{Simulation Design}
\label{sec:sim_design}

Each DGP follows Equation~\oldref{eq:dgp}, with geometric adstock and one of the four saturation forms (Hill C, Hill S, Weibull C, Weibull S). Two scenarios vary the number of channels and locations: S1 (4 channels, 5 locations) and S2 (6, 8). All use $156$ training weeks and $52$ test weeks. Adstock rates are drawn from $\text{Uniform}(0.5, 0.8)$; shape parameters from $\text{Uniform}(0.3, 0.6)$ for C-shape and $\text{Uniform}(2.0, 4.0)$ for S-shape DGPs; scale parameters $K_i$ (the half-saturation point for Hill and the 63.2\% point for Weibull) are linearly spaced from $8{,}000$ to $4{,}000$; channel mean coefficients are linearly spaced from $1.5$ to $0.3$, and location coefficients are drawn as $\mathcal{N}$(channel mean, $0.1^2$) and floored at $0.01$; location intercepts are $\mathcal{N}(0, 0.1^2)$; the other coefficients are $\beta_0=7.5$, $\beta_p=-1.2$, $\beta_d=0.015$, $\beta_u=-0.05$, $\beta_c=0.01$, $\beta_t=0.001$, $\beta_s^{(1)}=0.08$, $\beta_s^{(2)}=0.05$, and $\gamma=(0.10, 0.15, 0.20)$; and $\sigma = 0.02$. Weekly spend is drawn independently across weeks and locations from a multivariate normal distribution with pairwise channel correlation $\rho=0.3$, floored at \$100. With $\sigma=0.02$ the conditional coefficient of variation of sales is about 2\%, and the media term explains a median 86\% of the variance of log sales (Appendix Table~\oldref{tab:diagnostics}), so the baseline has a strong signal; the noise study raises $\sigma$ tenfold. \oldref{app:replicated} describes how the controls are generated and the spend support relative to $K_i$. \\

Each DGP is fitted with the four saturation candidates under two conditions. \emph{Adstock-included} candidates contain the carryover of the DGP, so exactly one candidate is correctly specified. \emph{Adstock-omitted} candidates drop adstock, so every candidate is wrong. In the simulations the estimator adds a mild Gaussian penalty that shrinks location media effects towards their channel mean. For each fit we compute out-of-sample sales metrics ($R^2$, MAPE, RMSE) and RoAS metrics (RoAS-MAE, RoAS-MAPE, RoAS-RMSE) against the known RoAS. The traditional criterion picks the highest-$R^2$ candidate; the benchmark criterion picks the lowest RoAS-MAE. With noiseless known RoAS the benchmark criterion is an oracle; it serves to measure how often, and by how much, fit-based ranking departs from RoAS ranking. All comparisons are within dataset, so they do not depend on the overall error level of the design. For the correctly specified candidate, the fitted penalized likelihood exceeds the likelihood at the generating parameters in all 40 diagnostic datasets, every start reports convergence, and RoAS computed at the generating parameters reproduces the known RoAS (Appendix Table~\oldref{tab:diagnostics}). The results are robust to local optima: in 14 of the 40 datasets a start at the generating parameters reaches a higher optimum than the five random starts, and 20 starts reduce this to 2 of 40 with no gain in RoAS recovery or agreement (Appendix Table~\oldref{tab:starts}). All candidates are fitted the same way, so the comparisons are like for like.

\subsection{Replicated Simulation Evidence}
\label{sec:replicated}

\ifsweeps We replicate the design and extend it in four directions;\else We replicate the design in a Monte Carlo study;\fi{} \oldref{app:replicated} gives cell-level results. Each replication redraws the DGP parameters, spend, controls, and noise, and re-estimates all four candidates.

\ifsweeps\begin{enumerate}
    \item \textbf{Monte Carlo replication.}\fi{} The main design is replicated $40$ times for each DGP, for S1 and S2, and for both adstock conditions ($320$ datasets per condition, $2{,}560$ model fits).
\ifsweeps
    \item \textbf{Media-spend collinearity.} The pairwise spend correlation is varied over $\rho \in \{0, 0.3, 0.6, 0.9\}$ (baseline $0.3$), holding everything else fixed and reusing the same random seeds across $\rho$ (S2, adstock-included candidates, $60$ datasets per value).
    \item \textbf{Noise level.} The error standard deviation is varied over $\sigma \in \{0.02, 0.05, 0.10, 0.20\}$ (baseline $0.02$), again with common random seeds (S2, adstock-included, $60$ datasets per value).
    \item \textbf{Adstock form.} Sales are generated with a delayed adstock that peaks one or two weeks after spend (kernel $w_k \propto \alpha^{(k-\theta)^2}$, $\theta \in \{1,2\}$), truncated at 13 lags and scaled to the same total carryover as geometric adstock. All candidates keep geometric adstock, so every candidate has the wrong carryover (S2, $60$ datasets).
\fi
\ifsweeps\end{enumerate}\fi

\ifsweeps In all four studies we also evaluate\else We also evaluate\fi{} selection with \emph{noisy} experiments. The analyst observes $b_i = \RoAS_i^{\text{true}}(1 + c\,z_i)$ with $z_i \sim \mathcal{N}(0,1)$ and a coefficient of variation $c \in \{0.10, 0.25, 0.50\}$ per channel, picks the candidate closest to $b$ in RoAS-MAE, and the pick is scored against the true RoAS (averaged over $500$ benchmark draws per dataset). This removes the oracle element and asks whether realistically imprecise experiments still beat fit-based selection. Table~\oldref{tab:robust_summary} summarizes the results; Appendix Figure~\oldref{fig:mc_ratio} shows the Monte Carlo distribution of the RoAS-error penalty from $R^2$ selection. The benchmark noise is unbiased, independent across channels, and available for every channel. Ratios are ratios of mean RoAS-MAE, which give more weight to settings with larger RoAS. Cell-level results are in \oldref{app:replicated}, and Appendix Table~\oldref{tab:selection_detail} gives paired comparisons, including the share of dataset--draw pairs in which the benchmark pick is better or materially worse.

{%
\begin{table}[htbp]
\centering
\caption{Predictive-Fit Selection versus RoAS Recovery Across Replicated Simulation Studies}
\label{tab:robust_summary}
\scriptsize
\resizebox{\textwidth}{!}{%
\begin{tabular}{@{}lccccccc@{}}
\toprule
& & \multicolumn{3}{c}{\textbf{$R^2$ selection vs.\ RoAS-best}} & \multicolumn{3}{c}{\textbf{RoAS-MAE relative to $R^2$ selection}} \\
\cmidrule(lr){3-5}\cmidrule(lr){6-8}
Study & Datasets & $\Pr(\text{agree})$ [95\% CI] & $\bar\tau$ & MAE ratio & Exp-10 & Exp-25 & Exp-50 \\
\midrule
Monte Carlo, S1, adstock-included & 160 & 0.71 [0.64, 0.78] & 0.83 & 1.22 & 1.00 & 1.56 & 2.48 \\
Monte Carlo, S2, adstock-included & 160 & 0.80 [0.73, 0.85] & 0.90 & 1.13 & 1.12 & 1.74 & 3.07 \\
Monte Carlo, S1, adstock-omitted & 160 & 0.29 [0.23, 0.37] & 0.14 & 1.05 & 0.95 & 0.96 & 0.97 \\
Monte Carlo, S2, adstock-omitted & 160 & 0.36 [0.29, 0.44] & 0.25 & 1.03 & 0.97 & 0.97 & 0.98 \\
\addlinespace
Collinearity, $\rho=0$ & 60 & 0.82 [0.70, 0.89] & 0.89 & 1.07 & 1.14 & 1.76 & 3.00 \\
Collinearity, $\rho=0.3$ & 60 & 0.85 [0.74, 0.92] & 0.92 & 1.09 & 1.17 & 1.78 & 3.21 \\
Collinearity, $\rho=0.6$ & 60 & 0.73 [0.61, 0.83] & 0.87 & 1.10 & 1.11 & 1.62 & 2.77 \\
Collinearity, $\rho=0.9$ & 60 & 0.75 [0.63, 0.84] & 0.83 & 1.08 & 1.02 & 1.38 & 2.42 \\
\addlinespace
Noise, $\sigma=0.02$ & 60 & 0.78 [0.66, 0.87] & 0.90 & 1.17 & 1.09 & 1.86 & 3.47 \\
Noise, $\sigma=0.05$ & 60 & 0.73 [0.61, 0.83] & 0.83 & 1.15 & 0.99 & 1.38 & 2.28 \\
Noise, $\sigma=0.1$ & 60 & 0.68 [0.56, 0.79] & 0.71 & 1.22 & 0.88 & 1.11 & 1.63 \\
Noise, $\sigma=0.2$ & 60 & 0.52 [0.39, 0.64] & 0.52 & 1.30 & 0.80 & 0.93 & 1.18 \\
\addlinespace
Adstock form, delayed-adstock DGP & 60 & 0.35 [0.24, 0.48] & 0.32 & 1.30 & 0.79 & 0.85 & 0.93 \\
\bottomrule
\end{tabular}}
\begin{tablenotes}\scriptsize
\item \textit{Notes}: Within each row, each dataset is an independent draw of generating parameters, media spend, controls and noise (the collinearity and noise studies reuse the same seeds across values of $\rho$ and $\sigma$); four candidate saturation specifications are fitted to each. $\Pr(\text{agree})$: share of datasets in which the highest test-$R^2$ candidate is also the lowest-RoAS-MAE candidate (Wilson 95\% interval; random choice among four candidates gives 0.25). $\bar\tau$: mean Kendall rank correlation between the $R^2$ ranking and the RoAS-accuracy ranking. MAE ratio: mean RoAS-MAE of the $R^2$-selected model divided by the mean RoAS-MAE of the RoAS-best model. Exp-$c$: mean RoAS-MAE (against the true RoAS) of the model selected against a \emph{noisy} benchmark with $c$\% coefficient of variation per channel, divided by the mean RoAS-MAE of the $R^2$-selected model; values below one favour benchmark-based selection. Collinearity and noise studies use S2 with adstock-included candidates; the adstock-form study uses S2 with a delayed-adstock DGP and geometric-adstock candidates.
\end{tablenotes}
\end{table}
}

\paragraph{Finding 1: With the correct settings and a strong signal, $R^2$ usually works, but not always} With adstock-included candidates, the highest-$R^2$ model is also the best RoAS model in $75.6\%$ of the $320$ datasets ($95\%$ CI $70.6$--$80.0\%$). Random choice would give $25\%$. The mean Kendall correlation between the two rankings is $0.86$, and out-of-sample MAPE gives the same picture (agreement $76.2\%$). So in this easy design fit is a useful guide, but not a reliable one. In the remaining 78 datasets $R^2$ picks another model, almost always by a tiny margin: the median $R^2$ difference between the pick and the best RoAS model is $0.00006$, and it is below $0.001$ in $98.7\%$ of them. These misses are costly. The mean RoAS-MAE of the $R^2$ pick is $1.15$ times that of the best RoAS model ($95\%$ interval $1.10$--$1.21$). When models fit almost equally well, as in Section~\oldref{sec:motivating}, a higher $R^2$ says little about which RoAS to trust.

\paragraph{Finding 2: When every candidate omits carryover, $R^2$ cannot rank them, and no selection removes the shared bias} With adstock-omitted candidates, agreement falls to $32.8\%$ ($27.9$--$38.1\%$), only a little above random choice, and the rank correlation to $0.19$. All four candidates share the same omission, so their RoAS errors are similar. The penalty from $R^2$ selection is small ($1.03$ times). Noisy benchmarks reduce RoAS-MAE by only $2$--$3\%$, although they find the best RoAS candidate much more often ($99\%$, $95\%$, and $83\%$ of selections at $c=0.10$, $0.25$, and $0.50$). The shared bias is large: the mean RoAS-MAE of the $R^2$ pick is $7.55$, against $0.42$ with adstock-included candidates. In this design out-of-sample $R^2$ reveals the omission ($0.995$ against $0.658$ for the best candidate of each set), and so does the benchmark. With $25\%$ benchmark noise, the best-fitting adstock-omitted model deviates from the benchmark by a median $77\%$ of benchmark RoAS, against $20\%$ for the best-fitting adstock-included model. A large gap to an experiment means the model needs to be revised. Choosing among the same wrong candidates cannot fix it.

\paragraph{Finding 3: Benchmarks beat $R^2$ where $R^2$ is weak, if they are precise} How much a benchmark helps depends on the setting. With adstock-included candidates in the baseline design, $R^2$ already selects well, and noisy benchmarks do worse than $R^2$: their mean RoAS-MAE is $1.09$, $1.69$, and $2.93$ times that of $R^2$ selection at benchmark coefficients of variation of $10\%$, $25\%$, and $50\%$. Where $R^2$ is weak, the order reverses (Findings 4 and 5). Benchmarks are most useful when sales are noisy or the carryover is wrong. An imprecise benchmark should not override a clear difference in fit.

\ifsweeps
\paragraph{Finding 4: Noise weakens selection on fit; collinearity matters less} As the noise standard deviation rises from $\sigma=0.02$ to $0.20$ (mean test $R^2$ of the best candidate from $0.996$ to $0.73$), agreement falls from $78.3\%$ to $51.7\%$ and the rank correlation from $0.90$ to $0.52$. At the same time the RoAS-MAE penalty from $R^2$ selection rises from $1.17$ to $1.30$ times. Benchmarks gain value accordingly. A $10\%$-noise benchmark cuts RoAS-MAE relative to $R^2$ selection by $12\%$ at $\sigma=0.10$ and by $20\%$ at $\sigma=0.20$, and a $25\%$-noise benchmark by $7\%$ at $\sigma=0.20$ (95\% interval $0.81$--$1.06$ for the ratio). Collinearity matters less in this design. As $\rho$ rises from $0$ to $0.9$, agreement moves from $81.7\%$ to $75.0\%$, the penalty stays between $1.07$ and $1.10$ times, and the $R^2$ range across candidates widens from $0.05$ to $0.10$. Noisy benchmarks do not beat $R^2$ at any $\rho$ (Appendix Figure~\oldref{fig:sweeps}).

\paragraph{Finding 5: A wrong carryover shape makes $R^2$ unreliable and benchmarks valuable} Here sales are generated with a delayed adstock that peaks one or two weeks after spend, while every candidate assumes geometric carryover. The highest-$R^2$ model is then the best RoAS model in only $35.0\%$ of datasets ($95\%$ CI $24.2$--$47.6\%$), the rank correlation is $0.32$, and the $R^2$ pick has $1.30$ times the RoAS-MAE of the RoAS-best model ($1.19$--$1.43$; Appendix Table~\oldref{tab:adstock_cells}). Omitting carryover makes all candidates about equally wrong (Finding 2); getting its shape wrong does not. The candidates' RoAS errors stay far enough apart for the choice to matter, and noisy benchmarks reduce RoAS-MAE relative to $R^2$ selection by $21\%$, $15\%$, and $7\%$ at $c=0.10$, $0.25$, and $0.50$ (95\% interval for the last ratio $0.85$--$1.02$). This is the concern raised by \cite{palu2026adstock}: carryover specifications that fit about equally well can imply different returns. In real data the true carryover shape is never known, so this case is a realistic one.
\fi

\subsection{Discussion}

The simulations show when the problem in the motivating example arises. Fit is a reasonable guide to RoAS when the candidate set contains the correct settings and the sales signal is strong. It becomes unreliable when the models fit almost equally well, when sales are noisy, or when all candidates share the same misspecification. In real data the analyst does not know which of these cases applies, because the correct adstock and saturation are unknown and the noise level is only partly visible. Experiments are most valuable in the hard cases, and how much they help depends on how precise they are. Two practical points follow. RoAS-MAE is on the original RoAS scale, so comparisons should be made within a dataset, not across scenarios. And because the value of an experiment depends on its precision and on how far all candidates are from the truth, the experimental uncertainty should be reported with the calibration result.

\subsection{When No Benchmark Is Available: RoAS Sensitivity and Consensus Diagnostics}
\label{sec:scenario2}

Without a benchmark, there is no way to tell which specification has the best RoAS, and fit metrics do not help. What the analyst can do is look at how RoAS varies across plausible specifications. A simple summary is the componentwise median RoAS vector: we pick the specification whose RoAS is closest to it, and use the known RoAS only afterwards to score the choice. The simulations show why it should not be used to select a model. With adstock-included candidates the median-consensus model is much less accurate than the $R^2$ pick (mean RoAS-MAE $3.59$ times; better in $7\%$ of datasets), because the median of the four RoAS vectors is often far from the correct model. With adstock-omitted candidates and under a wrong carryover shape, it is about as accurate as $R^2$ selection ($1.02$ and $1.01$ times). Its choice also depends on conventions, such as the upper or midpoint median and dropping a single candidate (Appendix Table~\oldref{tab:selection_detail}). A consensus summary cannot replace causal validation. Without a benchmark, analysts should report the median, interquartile range, and full range of RoAS across plausible specifications, and present them as a sensitivity statement rather than a causal estimate.

\section{Application: Two Advertisers with Geo-Holdout Experiments}
\label{sec:real}

We now apply the framework to real data from two advertisers. As in the simulations, several plausible specifications fit about equally well but give different RoAS. Unlike in the simulations, the true RoAS is unknown, and geo-holdout experiments provide the external check.

\subsection{Data and Business Context}

We use weekly national data for two advertisers (Datasets 1 and 2). The data come from the input of a production MMM pipeline and are used with permission. Client and brand identities are withheld, and channels have generic names. Each dataset has 156 weeks of conversions, the outcome the experiments measure. It also has weekly spend for every paid channel in the production model (10 channels in Dataset 1 and 6 in Dataset 2) and that model's non-media controls and event indicators (3 and 5). The benchmarks come from completed geo-holdout experiments in which a channel was turned off in test regions for 5 to 12 weeks: 16 experiments on four channels in Dataset 1 and 8 experiments on two channels in Dataset 2. Experiments on the same channel are pooled with inverse-variance weights. The pooled estimates have coefficients of variation of 0.11--0.19 in Dataset 1 and 0.05--0.07 in Dataset 2, the precise end of the range studied in Section~\oldref{sec:replicated}. RoAS is measured in incremental conversions per \$1{,}000 of spend.

\subsection{Model Specifications and Estimation}

We fit four specifications (Model 1 = Hill C-shape, Model 2 = Hill S-shape, Model 3 = Weibull C-shape, Model 4 = Weibull S-shape). They share the same geometric adstock, channels, and controls, and differ only in the saturation form. Each is estimated by constrained maximum likelihood as in Section~\oldref{sec:method}, with a single location, nonnegative media coefficients, and 20 random starts. For estimation, each channel's spend is rescaled to a mean of 5{,}000, and RoAS is converted back to the original units. History is short and the goal is RoAS, not forecasting, so we estimate on the full sample, as is common in practice; fit is therefore in-sample $R^2$ on the conversions scale. To match the experiments, the model RoAS for each experiment is computed over that experiment's test weeks (Equation~\oldref{eq:RoAS}, with the channel set to zero in those weeks only). The results are then pooled across a channel's experiments with the same weights as the experimental estimates.

\subsection{Results and Model Comparison}

Table~\oldref{tab:app_results} gives each model's RoAS for the benchmarked channels, with the pooled benchmarks and in-sample $R^2$.

{%
\begin{table}[htbp]
\centering
\caption{Application: RoAS by Specification and Experimental Benchmarks}
\label{tab:app_results}
\resizebox{\textwidth}{!}{%
\begin{tabular}{@{}llccccc@{}}
\toprule
\multicolumn{7}{@{}l}{\textit{Dataset 1} (156 weeks, 10 channels, 16 experiments)} \\
\textbf{Model} & \textbf{Saturation} & In-sample $R^2$ & \textbf{Search 1} & \textbf{Social} & \textbf{Online video} & \textbf{Display 2} \\
\midrule
Model 1 & Hill C & 0.979 & 1.9 & 4.6 & 0.0 & 32.8 \\
Model 2 & Hill S & 0.985 & 1.1 & 0.1 & 0.0 & 59.2 \\
Model 3 & Weibull C & 0.983 & 2.4 & 0.8 & 0.0 & 35.9 \\
Model 4 & Weibull S & \textbf{0.986} & 0.5 & 0.1 & 0.0 & 32.9 \\
\textbf{Benchmark} (SE) & --- & --- & \textbf{2.6} (0.4) & \textbf{0.9} (0.1) & \textbf{11.8} (1.2) & \textbf{11.4} (2.2) \\
Range across models & --- & --- & 0.5--2.4 & 0.1--4.6 & 0.0--0.0 & 32.8--59.2 \\
\midrule
\multicolumn{7}{@{}l}{\textit{Dataset 2} (156 weeks, 6 channels, 8 experiments)} \\
\textbf{Model} & \textbf{Saturation} & In-sample $R^2$ & \textbf{Search 2} & \textbf{Social} &  &  \\
\midrule
Model 1 & Hill C & 0.752 & 17.6 & 3.0 &  &  \\
Model 2 & Hill S & \textbf{0.774} & 0.9 & 34.3 &  &  \\
Model 3 & Weibull C & 0.753 & 18.7 & 11.9 &  &  \\
Model 4 & Weibull S & 0.770 & 0.1 & 6.7 &  &  \\
\textbf{Benchmark} (SE) & --- & --- & \textbf{8.1} (0.4) & \textbf{5.7} (0.4) &  &  \\
Range across models & --- & --- & 0.1--18.7 & 3.0--34.3 &  &  \\
\bottomrule
\end{tabular}%
}
\begin{tablenotes}
\small
\item \textit{Notes}: RoAS is incremental conversions per \$1{,}000 of spend. Model RoAS is computed over the test weeks of each experiment (Equation~\oldref{eq:RoAS}, the channel set to zero in those weeks only) and pooled across the experiments on a channel with the same inverse-variance weights as the benchmark. Benchmarks are inverse-variance-weighted means of geo-holdout estimates. Models are estimated on the full sample; $R^2$ is in-sample on the conversions scale (highest in bold).
\end{tablenotes}
\end{table}
}

\subsection{Model Selection and RoAS Recovery}

Table~\oldref{tab:app_recovery} reports recovery metrics for each specification and the picks of three criteria: highest $R^2$, lowest benchmark RoAS-MAPE, and the median heuristic of Section~\oldref{sec:scenario2}. \\

{%
\begin{table}[htbp]
\centering
\caption{Agreement with Benchmark RoAS}
\label{tab:app_recovery}
\resizebox{\textwidth}{!}{%
\begin{tabular}{@{}lcccc@{}}
\toprule
\textbf{Model Specification} & \textbf{RoAS-MAPE} & \textbf{RoAS-MAE} & \textbf{RoAS-RMSE} & In-sample $R^2$ \\
\midrule
\multicolumn{5}{@{}l}{\textit{Dataset 1}} \\
Model 1 (Hill C-shape) & 180.0\% & 9.36 & 12.32 & 0.979 \\
Model 2 (Hill S-shape) & 167.0\% & 15.48 & 24.63 & 0.985 \\
Model 3 (Weibull C-shape) & 84.6\% & 9.16 & 13.58 & 0.983 \\
Model 4 (Weibull S-shape) & 115.2\% & 9.04 & 12.29 & \textbf{0.986} \\
\quad Best by $R^2$ & \multicolumn{4}{l}{Model 4 (RoAS-MAPE 115.2\%)} \\
\quad Best by benchmark RoAS-MAPE & \multicolumn{4}{l}{Model 3 (RoAS-MAPE 84.6\%)} \\
\quad Median-consensus selection & \multicolumn{4}{l}{Model 3 (RoAS-MAPE 84.6\%)} \\
\midrule
\multicolumn{5}{@{}l}{\textit{Dataset 2}} \\
Model 1 (Hill C-shape) & 81.8\% & 6.07 & 6.98 & 0.752 \\
Model 2 (Hill S-shape) & 297.9\% & 17.94 & 20.91 & \textbf{0.774} \\
Model 3 (Weibull C-shape) & 120.0\% & 8.39 & 8.66 & 0.753 \\
Model 4 (Weibull S-shape) & 58.9\% & 4.55 & 5.71 & 0.770 \\
\quad Best by $R^2$ & \multicolumn{4}{l}{Model 2 (RoAS-MAPE 297.9\%)} \\
\quad Best by benchmark RoAS-MAPE & \multicolumn{4}{l}{Model 4 (RoAS-MAPE 58.9\%)} \\
\quad Median-consensus selection & \multicolumn{4}{l}{Model 4 (RoAS-MAPE 58.9\%)} \\
\bottomrule
\end{tabular}%
}
\begin{tablenotes}
\small
\item \textit{Notes}: Errors are computed against the pooled benchmark RoAS of the tested channels (equal channel weights; RoAS per \$1{,}000). Median-consensus selection picks the specification closest (mean absolute distance, each channel scaled by its mean absolute RoAS across specifications) to the componentwise median RoAS vector; ties go to the lowest model number.
\end{tablenotes}
\end{table}
}

The results look like the harder simulation settings. In both datasets the four specifications fit almost equally well (in-sample $R^2$ of 0.979--0.986 in Dataset 1 and 0.752--0.774 in Dataset 2), yet their RoAS differ widely. Display 2 ranges from 32.8 to 59.2 in Dataset 1, and Social from 3.0 to 34.3 in Dataset 2. The model with the highest $R^2$ is not the one closest to the experiments. In Dataset 1, Model 4 has the highest $R^2$ (0.986) and a RoAS-MAPE of 115.2\%, against 84.6\% for Model 3. The $R^2$ pick thus has 1.36$\times$ the benchmark discrepancy. By RoAS-MAE, which gives more weight to high-RoAS channels, the two are close, with a ratio of 0.99. In Dataset 2, Model 2 has the highest $R^2$ (0.774) and a RoAS-MAPE of 297.9\%, 5.06$\times$ that of Model 4 (58.9\%), whose $R^2$ is almost the same (0.770); the RoAS-MAE ratio is 3.95. The median diagnostic picks the model closest to the experiments in Dataset 1 under both the midpoint and the upper median, and in Dataset 2 under the midpoint median. As in the simulations, its choice depends on conventions: dropping one of the three non-selected candidates changes its choice in 1 case in each dataset.

Choosing among saturation forms does not remove the error that all candidates share. Even the benchmark-closest specification misses the experiments by a RoAS-MAPE of 84.6\% and 58.9\%. In Dataset 1, every candidate gives Online video zero RoAS, and all four overstate Display 2 RoAS by 186\% to 418\%. This is the pattern of Finding 2. When all candidates share a misspecification, here plausibly in how conversions are split among channels whose spend moves together, no choice among them fixes it, and the comparison with the experiments is what reveals it. This is also why fit cannot separate the saturation forms: shape parameters lie at a bound in 8 to 13 of the 30 in Dataset 1 and in 1 to 10 of the 18 in Dataset 2, depending on the candidate, so sales data alone pin down the curvature only loosely, and the experiments supply the information needed to choose. In both datasets, calibrating against the experiments changes which specification would be used for budget decisions.

\section{Managerial Implications}
\label{sec:implications}

Good predictive fit is not enough for media investment decisions. A model can fit past sales well and still give credit to the wrong channels. In real data the correct adstock and saturation are unknown, so a practitioner cannot rely on $R^2$ alone to pick the model. When experiments exist, use them to choose among specifications, after checking that they measure the same RoAS as the model. Our simulations show that experiments help most when sales are noisy or the carryover is wrong, and that they help less as they become less precise. An imprecise experiment should qualify a clear difference in fit, not override it. Without experiments, report RoAS across plausible specifications instead of relying on one model chosen by fit. Report forecast checks and causal checks separately: $R^2$, MAPE, and RMSE are useful, but they are not evidence that the RoAS is right.

\section{Concluding Remarks} \label{sec:conclusions}
Good predictive fit does not mean good RoAS recovery in MMM. Across hundreds of simulated datasets, the best-fitting model was not the one closest to the true RoAS in about one dataset in four when the model settings were correct and the signal was strong, and in about half or more when sales were noisy or the carryover was wrong. In real data the analyst cannot tell which of these cases applies, so the fit alone cannot be trusted to pick the model. \\

Evaluation criteria should match the purpose of the model. $R^2$, MAPE, and RMSE are the right tools for forecasting. For RoAS, candidate specifications should be compared with experimental estimates of the same quantity. Without experiments, fit cannot identify the specification with the best RoAS, and analysts should report how RoAS changes across plausible specifications. \\

Our simulations cover a wide range of response curves, model sizes, noise levels, spend correlations, and carryover forms, and the two real applications show the same pattern. The approach is not tied to a particular model. Our candidates use geometric carryover, but richer forms, such as the rational transfer functions advocated by \cite{palu2026adstock}, can be compared against experiments in the same way. We focus on average RoAS over the observed spend path, which is what geo-holdout experiments measure; with experiments at different spend levels, the same approach applies to marginal RoAS. Like all MMMs, our models rely on assumptions such as exogenous spend and no unmodeled confounding, and calibration against experiments checks the RoAS without relying on them. Natural next steps include larger channel sets, experiments that cover only some channels, endogenous spend, precision-weighted losses, channel interactions, and Bayesian calibration against experiments. \\

\vspace{1em}
\noindent \textbf{Data and Code Availability:} The Python code (including the replication code for all simulation studies) and the simulated datasets are available on request. The anonymized data analysed in Section~\oldref{sec:real} are provided with the code; client and brand identities are withheld.

\vspace{1em}
\noindent \textbf{Disclosure of interest:} The authors declare no competing interests. The views expressed are the authors' own and do not represent those of their affiliated organizations.\

\vspace{1em}
\noindent \textbf{Declaration of funding:} This research received no external funding.

\newpage

\clearpage
\appendix
\renewcommand{\thetable}{\Alph{section}.\arabic{table}}
\renewcommand{\thefigure}{\Alph{section}.\arabic{figure}}

\section{Details of the Motivating Example}
\label{app:motivating}
\setcounter{table}{0}\setcounter{figure}{0}
Table~\oldref{tab:variables} lists the variables of the data-generating process in Equation~\oldref{eq:dgp}, Table~\oldref{tab:true_params} gives the media parameters of the example dataset (Monte Carlo seed 12014), and Figure~\oldref{fig:sales_metrics} compares the four candidates on out-of-sample sales metrics. The other coefficients are given in Section~\oldref{sec:sim_design}.

\begin{table}[htbp]
\centering
\caption{Variable Definitions and Descriptions}
\label{tab:variables}
\resizebox{\textwidth}{!}{%
\begin{tabular}{@{}llcp{3.2cm}@{}}
\toprule
\textbf{Variable} & \textbf{Description} & \textbf{Type} & \textbf{Dimension} \\
\midrule
\multicolumn{4}{@{}l}{\textit{Response Variable}} \\
$y_{lt}$ & Sales units & Response & $L \times T$ \\[0.5em]
\multicolumn{4}{@{}l}{\textit{Media Variables}} \\
$x_{i,lt}$ & Media spend in channel $i$ (dollars) & Predictor & $C \times L \times T$ \\
$m_{i,lt}$ & Transformed media (adstock + saturation) & Derived & $C \times L \times T$ \\[0.5em]
\multicolumn{4}{@{}l}{\textit{Pricing \& Promo Variables}} \\
$p_{lt}$ & Price (dollars) & Predictor & $L \times T$ \\
$d_{lt}$ & Discount percentage & Predictor & $L \times T$ \\[0.5em]
\multicolumn{4}{@{}l}{\textit{Macroeconomic Variables}} \\
$u_t$ & Unemployment rate & Predictor & $T$ \\
$c_t$ & Consumer Confidence Index & Predictor & $T$ \\[0.5em]
\multicolumn{4}{@{}l}{\textit{Holiday Indicators}} \\
$h_{1,t}$ & Independence Day indicator & Predictor & $T$ \\
$h_{2,t}$ & Thanksgiving indicator & Predictor & $T$ \\
$h_{3,t}$ & Christmas indicator & Predictor & $T$ \\[0.5em]
\multicolumn{4}{@{}l}{\textit{Seasonality, Trend and Location Variables}} \\
$t$ & Time trend (weeks) & Predictor & $T$ \\
$\sin\left(\frac{2\pi t}{52}\right)$ & Annual seasonality (sine component) & Predictor & $T$ \\
$\cos\left(\frac{2\pi t}{52}\right)$ & Annual seasonality (cosine component) & Predictor & $T$ \\
$\ind{l=j}$ & Location $j$ indicator & Predictor & $L$ \\
\bottomrule
\end{tabular}}
\end{table}

{%
\begin{table}[htbp]
\centering
\caption{Generating Media Parameters of the Example Dataset}
\label{tab:true_params}
\small
\resizebox{\textwidth}{!}{%
\begin{tabular}{lcccccc}
\toprule
Parameter & Channel 1 & Channel 2 & Channel 3 & Channel 4 & Channel 5 & Channel 6 \\
\midrule
Mean media coefficient $\bar b_i$ & 1.448 & 1.225 & 1.018 & 0.792 & 0.570 & 0.295 \\
Adstock decay $\alpha_i$ & 0.513 & 0.509 & 0.745 & 0.787 & 0.732 & 0.642 \\
Scale $K_i$ & 8,000 & 7,200 & 6,400 & 5,600 & 4,800 & 4,000 \\
Shape $O_i$ & 0.523 & 0.528 & 0.561 & 0.450 & 0.526 & 0.448 \\
Known RoAS (test weeks) & 4.420 & 4.484 & 4.328 & 3.239 & 3.191 & 1.973 \\
\bottomrule
\end{tabular}}
\begin{tablenotes}\footnotesize
\item \textit{Notes}: Monte Carlo dataset with seed 12014 (Hill C, S2: 6 channels, 8 locations). Other coefficients are as in Section~\oldref{sec:performance}.
\end{tablenotes}
\end{table}
}

\begin{figure}[htbp]
    \centering
    \includegraphics[width=\textwidth]{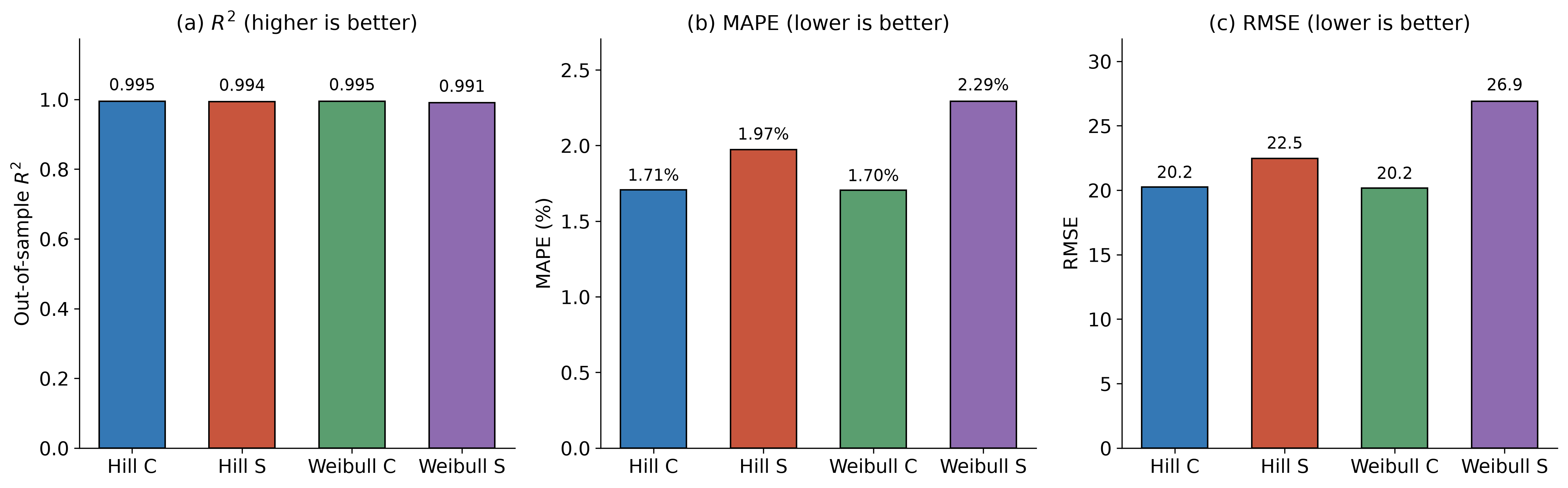}
    \caption{Sales Prediction Accuracy of the Four Candidates in the Example Dataset (test weeks). The Weibull C and Hill C candidates are practically indistinguishable on all three metrics.}
    \label{fig:sales_metrics}
\end{figure}

\clearpage
\section{Saturation Functions and Estimation Details}
\label{app:method}
\setcounter{table}{0}\setcounter{figure}{0}

\subsection{Hill and Weibull Saturation Families}
\paragraph{Hill Saturation Functions}

The Hill function, which comes from biochemistry, has two parameters:

\begin{equation}
\label{eq:hill}
f_{\text{Hill}}(z; K, O) = \frac{z^O}{K^O + z^O}
\end{equation}

where $K > 0$ is the half-saturation point (i.e., $f_{\text{Hill}}(K; K, O) = 0.5$) and $O > 0$ controls the curve shape.

\begin{itemize}
    \item \textbf{Hill S-shape} ($O > 1$): Sigmoidal saturation curve
    \begin{equation}
    \label{eq:hill_s}
    f_{\text{Hill-S}}(z; K, O) = \frac{z^O}{K^O + z^O}, \quad O > 1
    \end{equation}
    Returns first rise, then saturate; a larger $O$ makes the transition sharper. This shape is often used for brand campaigns that need a threshold level of awareness.
    
    \item \textbf{Hill C-shape} ($0 < O \leq 1$): C-shaped saturation curve
    \begin{equation}
    \label{eq:hill_c}
    f_{\text{Hill-C}}(z; K, O) = \frac{z^O}{K^O + z^O}, \quad 0 < O \leq 1
    \end{equation}
    Returns are highest at first and fall steadily; a smaller $O$ means faster saturation. This shape is typical of direct-response channels.
\end{itemize}

For $0 < O \leq 1$ the Hill curve is concave everywhere. For $O > 1$ it is sigmoidal, convex at low spend and concave at high spend, with inflection point
\[
z_{\mathrm{inf}}^{\mathrm{Hill}} = K\left(\frac{O-1}{O+1}\right)^{1/O}.
\]

\paragraph{Weibull Saturation Functions}

The Weibull distribution function, familiar from reliability and survival analysis, gives an alternative form:

\begin{equation}
\label{eq:weibull}
f_{\text{Weibull}}(z; K, O) = 1 - \exp\left(-\left(\frac{z}{K}\right)^O\right)
\end{equation}

where $K > 0$ is the scale parameter (the point at which $f_{\text{Weibull}}(K; K, O) \approx 0.632$) and $O > 0$ controls the shape.

\begin{itemize}
    \item \textbf{Weibull S-shape} ($O > 1$): S-shaped saturation curve
    \begin{equation}
    \label{eq:weibull_s}
    f_{\text{Weibull-S}}(z; K, O) = 1 - \exp\left(-\left(\frac{z}{K}\right)^O\right), \quad O > 1
    \end{equation}
    Response is slow at first, then accelerates, then saturates; a larger $O$ sharpens the transition. It offers a different functional form from the Hill S-shape for threshold-type response.
    
    \item \textbf{Weibull C-shape} ($0 < O \leq 1$): Concave saturation curve
    \begin{equation}
    \label{eq:weibull_c}
    f_{\text{Weibull-C}}(z; K, O) = 1 - \exp\left(-\left(\frac{z}{K}\right)^O\right), \quad 0 < O \leq 1
    \end{equation}
    As for the Hill C-shape, returns are highest at first and fall steadily; a smaller $O$ means faster saturation. It is often used for mature channels such as search.
\end{itemize}

For $0 < O \leq 1$ the Weibull curve is concave everywhere. For $O > 1$ it is sigmoidal, with inflection point
\[
z_{\mathrm{inf}}^{\mathrm{Weibull}} = K\left(\frac{O-1}{O}\right)^{1/O}.
\]

\paragraph{Comparative Properties}

Both families can be concave or sigmoidal, but they differ in four ways:

\begin{itemize}
    \item \textbf{Form}: the Hill function is rational, the Weibull function exponential.
    \item \textbf{Scale}: $f_{\text{Hill}}(K;K,O)=0.5$, whereas $f_{\text{Weibull}}(K;K,O)=1-e^{-1}\approx 0.632$.
    \item \textbf{Inflection points}: for $O>1$ the two sigmoidal curves bend at different points, given above.
    \item \textbf{Low spend}: the families can differ markedly near zero spend, and so imply different marginal RoAS there.
\end{itemize}

The choice between them should be guided by the data and, where available, by agreement with RoAS benchmarks that measure the same quantity. Figure~\oldref{fig:saturation_functions} in Section~\oldref{sec:method} shows the four functions.

\subsection{Likelihood and Constrained Estimation}
We estimate the model in Equation~\oldref{eq:main_model} by bound-constrained maximum likelihood. Let
\[
\tilde y_{lt} \equiv \log(y_{lt}),
\]
and define the linear predictor
\begin{equation}
\label{eq:eta}
\eta_{lt}(\Theta)
=
\beta_0
+
\sum_{j=2}^{L} \beta_{L,j}\,\ind{l=j}
+
\sum_{i=1}^{C}
\left(
  \beta_i + \sum_{j=2}^{L} \beta_{i,j}\,\ind{l=j}
\right) m_{i,lt}
+
\boldsymbol{X}_{lt}'\boldsymbol{\gamma},
\end{equation}
where $m_{i,lt} = f_{\mathrm{sat}}(g_{\mathrm{adstock}}(x_{i,lt};\alpha_i);K_i,O_i)$.

We assume
\[
\tilde y_{lt} = \eta_{lt}(\Theta) + \varepsilon_{lt},
\qquad
\varepsilon_{lt} \stackrel{\mathrm{iid}}{\sim} \mathcal{N}(0,\sigma^2).
\]
Hence, the conditional density of $\tilde y_{lt}$ given $\Theta$ is
\begin{equation}
\label{eq:obs_likelihood}
f(\tilde y_{lt}\mid \Theta)
=
\frac{1}{\sqrt{2\pi}\sigma}
\exp\!\left\{
-\frac{\bigl(\tilde y_{lt}-\eta_{lt}(\Theta)\bigr)^2}{2\sigma^2}
\right\}.
\end{equation}

The sample log-likelihood is therefore
\begin{equation}
\label{eq:loglik}
\ell(\Theta)
=
-\frac{LT}{2}\log(2\pi)
-
LT\log\sigma
-
\frac{1}{2\sigma^2}
\sum_{l=1}^{L}\sum_{t=1}^{T}
\bigl(\tilde y_{lt}-\eta_{lt}(\Theta)\bigr)^2.
\end{equation}

We maximize $\ell(\Theta)$ subject to economically motivated sign and shape constraints:
\begin{equation}
\label{eq:ml_constrained}
\begin{array}{rll}
\displaystyle \max_{\Theta} & \ell(\Theta) \\ [5pt]
\text{subject to} & \beta_i \geq 0 & i=1,\ldots,C \\ [3pt]
& \beta_i + \beta_{i,j} \geq 0 & i=1,\ldots,C,\; j=2,\ldots,L \\ [3pt]
& 0 \leq \alpha_i < 1 & i=1,\ldots,C \\ [3pt]
& K_i > 0,\; \rev{0<O_i\leq 1 \text{ (C-shape) or } O_i>1 \text{ (S-shape)},\; \sigma>0} & i=1,\ldots,C.
\end{array}
\end{equation}

The problem is non-convex because of the recursive adstock and the saturation functions, so we use multiple starts and keep the solution with the highest objective. Implementation details:
\begin{itemize}
    \item \textbf{Parameterization.} The total location-specific media effects $b_{i,l}=\beta_i+\beta_{i,l}$ (with $\beta_{i,1}=0$) are the free parameters, each bounded below by zero, so the nonnegativity constraints hold by construction. Spend enters in dollars without standardization, so $K_i$ is on the dollar scale.
    \item \textbf{Penalty.} In the simulations the objective is $\ell(\Theta)-\frac{1}{2\kappa^2}\sum_{i}\sum_{l}(b_{i,l}-\bar b_i)^2$ with $\kappa^2=1$, where $\bar b_i$ is the mean of $b_{i,\cdot}$ over locations; with a single location (Section~\oldref{sec:real}) the penalty is zero.
    \item \textbf{Bounds.} $\alpha_i\in[0.01,0.99]$, $K_i\in[100,20000]$, $O_i\in[0.1,0.99]$ for C-shape and $O_i\in[1.01,5]$ for S-shape candidates, $\sigma\geq 0.01$, nonnegative discount and holiday coefficients; other coefficients are unbounded.
    \item \textbf{Profiling.} For fixed $(\alpha,K,O,\sigma)$ the objective is a penalized least-squares problem in the remaining coefficients (intercepts, $b_{i,l}$, controls, holidays) with nonnegativity bounds; it is solved exactly (normal equations, or bounded-variable least squares when a bound binds). The numerical search is therefore over $(\alpha,K,O,\sigma)$ only, with $K_i$ searched in thousands of dollars. This changes the computation, not the estimator.
    \item \textbf{Starts.} Five random starts with $\alpha_i\sim U(0.2,0.7)$, $K_i\sim U(3000,10000)$, $O_i\sim U(0.4,0.9)$ (C-shape) or $U(1.5,3.5)$ (S-shape), and $\sigma=0.05$. \rev{The application (Section~\oldref{sec:real}) uses 20 starts from the same distribution, with each channel's spend rescaled to a mean of 5{,}000; Appendix Table~\oldref{tab:starts} repeats the diagnostics and the S1 Monte Carlo study with 20 starts.}
    \item \textbf{Optimizer.} L-BFGS-B over the profiled objective with finite-difference gradients, relative function tolerance $10^{-6}$, projected-gradient tolerance $10^{-8}$, at most 1{,}000 iterations per start. For correctly specified candidates the attained objective is compared with the objective at the generating parameters (\oldref{app:replicated}).
    \item \textbf{Failure policy.} A start that raises a numerical error is skipped; the best start is retained whether or not the optimizer reports convergence. Convergence flags, objective values for every start, and box-bound hits are stored for every fit; convergence and bound hits are summarized in \oldref{app:replicated}. A dataset in which a candidate cannot be fitted at all would be excluded from the summaries.
\end{itemize}

\clearpage
\section{Replicated Simulation Studies: Cell-Level Results}
\label{app:replicated}
\setcounter{table}{0}\setcounter{figure}{0}

\rev{This appendix reports the design, the estimation diagnostics, and the cell-level results of the replicated studies summarized in Section~\oldref{sec:replicated}. All studies use the candidate models, estimator (\oldref{app:method}), $156/52$-week split, and RoAS definition of Section~\oldref{sec:sim_design}. Controls are generated as $p_{lt}=\max(50+5z_{lt}+2\sin(2\pi t/52),\,30)$, $d_{lt}=0.3\,B_{lt}$ with $B_{lt}\sim\mathrm{Beta}(2,5)$, $u_t=5+0.5z_t$, $c_t=100+5z'_t$, and holiday indicators in weeks 26, 47, and 52 of each year, with independent standard normal $z$. Weekly channel spend has means linearly spaced from 5{,}000 to 3{,}000, coefficient of variation 1, and pairwise correlation $\rho$, floored at 100. Adstock-included and adstock-omitted candidates are fitted to the same datasets; each study has its own seeds, listed in the replication code. Across the 4{,}720 fits no fit failed, and 4 of 23{,}600 optimizer starts did not report convergence. At least one adstock or saturation parameter lies at a box bound in 0.9\% of fits of the correctly specified Monte Carlo candidate and in 76\% of the fits of the other adstock-included Monte Carlo candidates.}

\begin{figure}[htbp]
    \centering
    \includegraphics[width=0.85\textwidth]{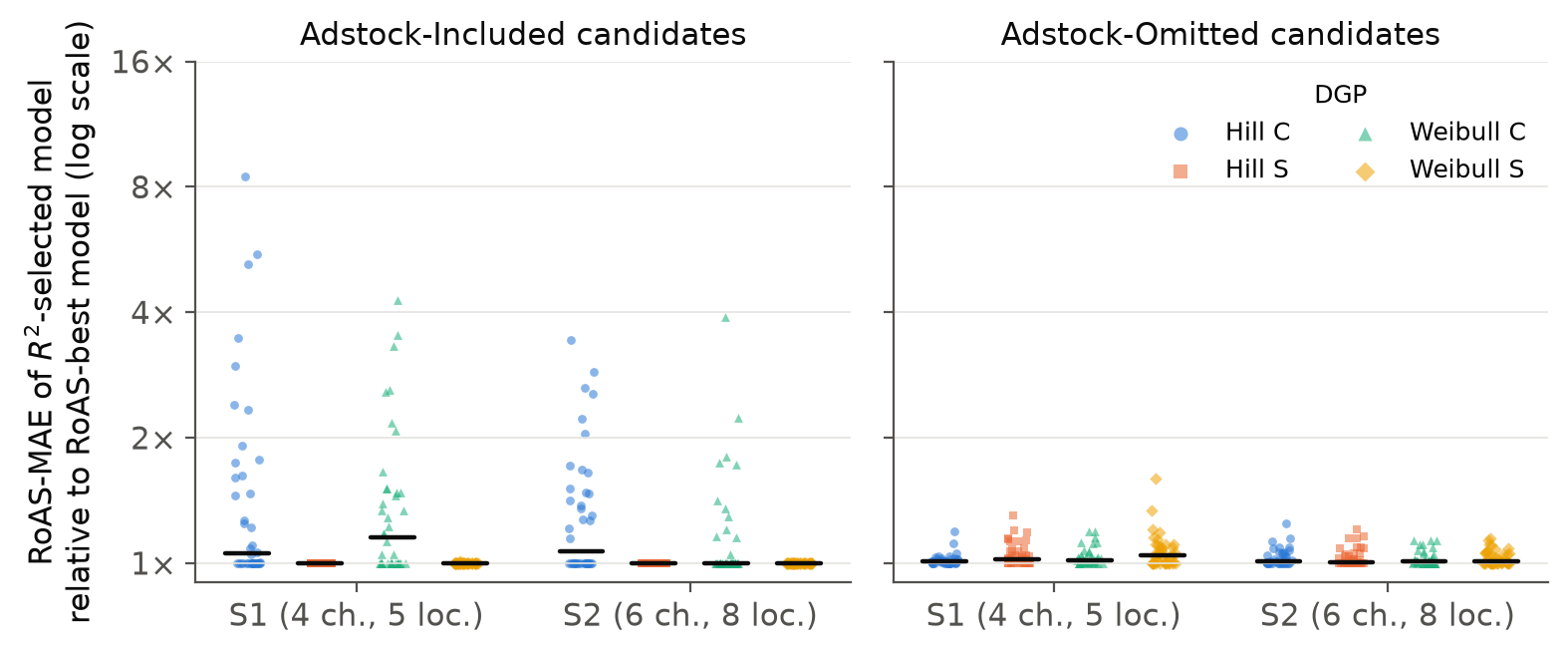}
    \caption{Monte Carlo distribution of the RoAS-error penalty from predictive-fit selection: RoAS-MAE of the highest-$R^2$ candidate divided by that of the RoAS-best candidate (log scale; $1\times$ means $R^2$ selected the RoAS-best model). Each point is one of 40 independent datasets per cell; black bars are medians.}
    \label{fig:mc_ratio}
\end{figure}

\begin{figure}[htbp]
    \centering
    \includegraphics[width=\textwidth]{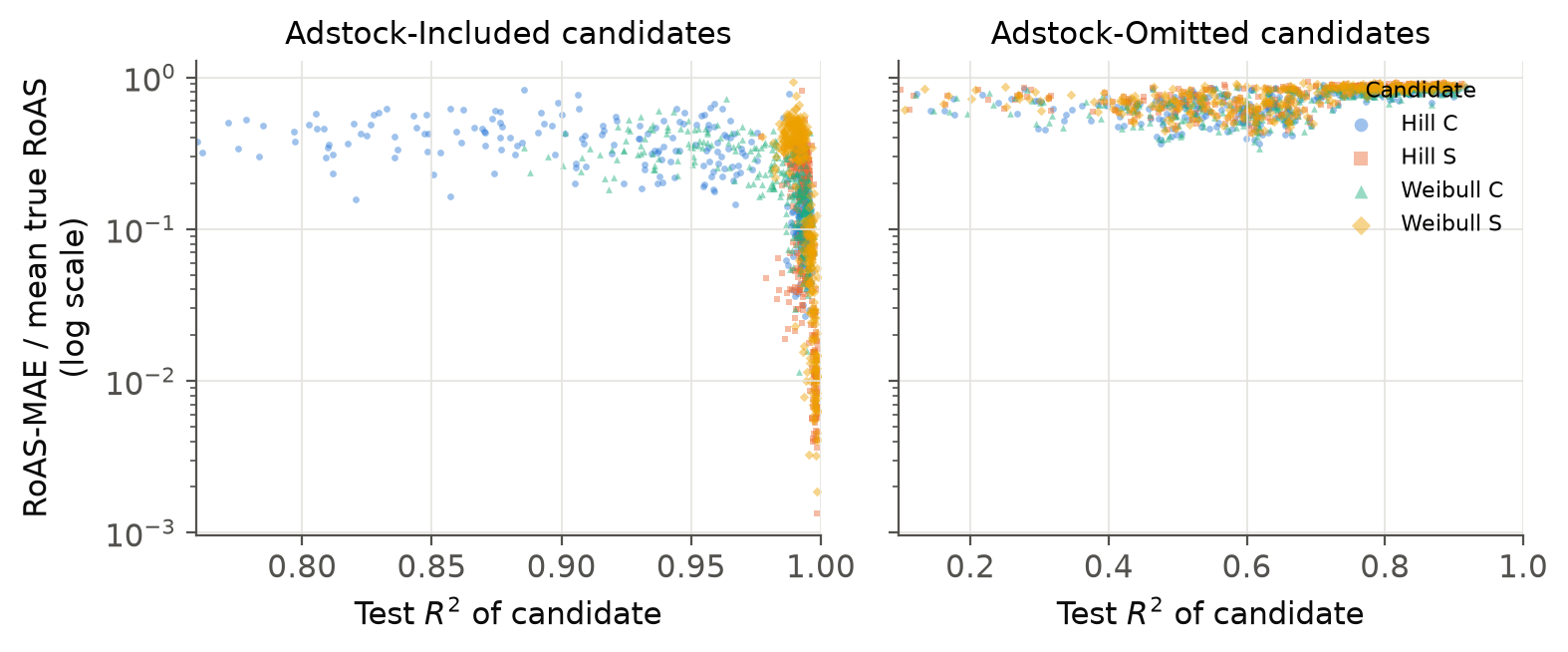}
    \caption{\rev{Out-of-sample $R^2$ against RoAS-MAE relative to mean true RoAS (log scale) for every fitted Monte Carlo candidate, by candidate specification.}}
    \label{fig:candidates}
\end{figure}

\rev{Table~\oldref{tab:design} summarizes the true RoAS and the spend support. Table~\oldref{tab:diagnostics} reports estimation diagnostics for the correctly specified candidate: in all 40 datasets the free fit attains a higher penalized likelihood than the generating parameter vector, the optimizer reports convergence for all starts, no shape parameter is at a bound, and the RoAS computed at the generating parameters equals the known RoAS. Fixing the adstock and saturation parameters at their generating values reduces the median RoAS-MAE from 0.34 to 0.08, and starting the optimizer at the generating values gives 0.15. In 14 of the 40 datasets that start reaches a log-likelihood more than one unit above the free fit (by up to 16.5), so the five random starts sometimes miss a better local optimum. The remaining RoAS error of the correct model therefore reflects mainly weak identification of the response-curve parameters in finite samples, not the optimizer stopping early; local optima matter little for RoAS. Table~\oldref{tab:starts} repeats the free fit with 20 random starts: the start at truth then finds a higher optimum in only 2 of the 40 datasets, but the median RoAS-MAE is essentially unchanged (0.339 against 0.335), and RoAS-MAE moves by more than 0.05 in both directions (lower in 14 datasets, higher in 12). Refitting all S1 Monte Carlo datasets with 20 starts changes the $R^2$ pick in 19\% of the adstock-included datasets but leaves the agreement rate unchanged (71\%); the MAE ratio of the $R^2$ pick rises from 1.22 to 1.29, and the 10\%-noise benchmark moves from parity with $R^2$ (1.00) to slightly better (0.97), while the 25\% and 50\% benchmarks remain worse. Figure~\oldref{fig:candidates} plots out-of-sample $R^2$ against relative RoAS error for every fitted Monte Carlo candidate. Table~\oldref{tab:mc_cells} and Figure~\oldref{fig:mc_ratio} report the Monte Carlo study by cell\ifsweeps, Tables~\oldref{tab:corr_cells}, \oldref{tab:noise_cells}, and~\oldref{tab:adstock_cells} report the stress tests by DGP, and Figure~\oldref{fig:sweeps} plots the agreement rate and the RoAS-error penalty of $R^2$ selection against $\rho$ and $\sigma$\fi{}. Table~\oldref{tab:selection_detail} gives paired comparisons for the noisy-benchmark and consensus rules.}

{%
\begin{table}[htbp]
\centering
\caption{Monte Carlo Design: True RoAS and Spend Support}
\label{tab:design}
\scriptsize
\begin{tabular}{@{}llccccc@{}}
\toprule
& & \multicolumn{2}{c}{\textbf{True RoAS (test weeks)}} & \multicolumn{2}{c}{\textbf{Spend / $K$}} & \textbf{Adstock / $K$} \\
\cmidrule(lr){3-4}\cmidrule(lr){5-6}\cmidrule(lr){7-7}
Scen. & DGP & Median & 5--95\% & Median & 5--95\% & 5--95\% \\
\midrule
S1 & Hill C & 1.33 & 0.65--1.76 & 0.67 & 0.01--1.80 & 0.79--3.94 \\
S1 & Hill S & 4.61 & 2.24--6.73 & 0.68 & 0.01--1.81 & 0.74--3.93 \\
S1 & Weibull C & 3.01 & 1.56--4.25 & 0.68 & 0.01--1.81 & 0.79--4.16 \\
S1 & Weibull S & 7.21 & 3.64--9.87 & 0.68 & 0.01--1.81 & 0.80--3.86 \\
\addlinespace
S2 & Hill C & 3.71 & 1.85--4.89 & 0.67 & 0.01--1.80 & 0.80--4.01 \\
S2 & Hill S & 20.48 & 11.03--33.42 & 0.67 & 0.01--1.79 & 0.78--3.88 \\
S2 & Weibull C & 10.85 & 6.08--15.48 & 0.67 & 0.01--1.79 & 0.76--3.74 \\
S2 & Weibull S & 39.28 & 20.85--54.12 & 0.68 & 0.01--1.80 & 0.77--3.94 \\
\addlinespace
\bottomrule
\end{tabular}
\begin{tablenotes}\scriptsize
\item \textit{Notes}: True RoAS pooled over channels and the 40 datasets per cell. Spend and adstocked spend are expressed relative to the channel scale parameter $K$ (median across datasets of the within-dataset quantiles). Weekly spend is floored at 100 (share of channel-weeks at the floor: 0.165), so the zero-spend counterfactual lies below the observed support.
\end{tablenotes}
\end{table}
}

{%
\begin{table}[htbp]
\centering
\caption{Estimation Diagnostics for the Correctly Specified Candidate (S2, adstock-included)}
\label{tab:diagnostics}
\scriptsize
\resizebox{\textwidth}{!}{%
\begin{tabular}{@{}lccccccccccc@{}}
\toprule
& & Identity & \multicolumn{3}{c}{Median RoAS-MAE} & \multicolumn{2}{c}{Free-fit objective vs.\ truth} & Start at truth & Converged & Shape param. & Media share \\
\cmidrule(lr){4-6}\cmidrule(lr){7-8}
DGP & $R$ & max rel.\ error & Free fit & Shapes fixed & Start at truth & Lower & Higher & higher by $>1$ & starts & at bound & of $\mathrm{Var}(\log y)$ \\
\midrule
Hill C & 10 & 0 & 0.309 & 0.048 & 0.087 & 1.00 & 0.00 & 0.30 & 1.00 & 0.00 & 0.58 \\
Hill S & 10 & 0 & 0.150 & 0.059 & 0.101 & 1.00 & 0.00 & 0.30 & 1.00 & 0.00 & 0.93 \\
Weibull C & 10 & 0 & 1.056 & 0.105 & 0.310 & 1.00 & 0.00 & 0.60 & 1.00 & 0.00 & 0.75 \\
Weibull S & 10 & 0 & 0.339 & 0.174 & 0.223 & 1.00 & 0.00 & 0.20 & 1.00 & 0.00 & 0.96 \\
\addlinespace
\textit{Pooled} & 40 & 0 & 0.339 & 0.084 & 0.149 & 1.00 & 0.00 & 0.35 & 1.00 & 0.00 & 0.86 \\
\bottomrule
\end{tabular}}
\begin{tablenotes}\scriptsize
\item \textit{Notes}: $R$ = datasets (the first ten Monte Carlo replications per DGP). Identity: largest relative difference between the RoAS computed by the estimation code at the generating parameters and the known RoAS. Free fit: the estimator used throughout (five random starts). Shapes fixed: adstock and saturation parameters fixed at their generating values, other parameters estimated. Start at truth: one start at the generating parameter vector. Lower / Higher: share of datasets in which the free-fit training objective is below / above the objective at the generating vector by more than $10^{-6}$ in relative terms. Start at truth higher by $>1$: share of datasets in which the fit started at the generating vector attains a log-likelihood more than one unit above the free fit (a better local optimum missed by the five random starts). Converged starts: mean share of starts for which the optimizer reports convergence. Shape param.\ at bound: share of free fits with at least one adstock or saturation parameter at a box bound. Media share: median share of the variance of log sales accounted for by the media term.
\end{tablenotes}
\end{table}
}

{%
\begin{table}[htbp]
\centering
\caption{Sensitivity to the Number of Random Starts}
\label{tab:starts}
\small
\resizebox{\textwidth}{!}{%
\begin{tabular}{@{}lcc@{}}
\toprule
& 5 starts & 20 starts \\
\midrule
\multicolumn{3}{@{}l}{\textit{A. Diagnostic datasets (S2, correctly specified candidate, $R=40$)}} \\
Start at truth higher by $>1$ (datasets) & 14 & 2 \\
20 starts higher than 5 starts by $>1$ (datasets) & --- & 24 \\
Median RoAS-MAE & 0.339 & 0.335 \\
Mean RoAS-MAE & 0.600 & 0.547 \\
RoAS-MAE lower / higher by $>0.05$ with 20 starts (datasets) & --- & 14 / 12 \\
\addlinespace
\multicolumn{3}{@{}l}{\textit{B. Monte Carlo, S1, adstock-included ($R=160$)}} \\
$\Pr(R^2\text{-best} = \text{RoAS-best})$ & 0.713 & 0.713 \\
MAE ratio, $R^2$ pick / RoAS-best & 1.22 & 1.29 \\
Mean RoAS-MAE of the $R^2$ pick & 0.202 & 0.182 \\
Exp-10 / Exp-25 / Exp-50 vs.\ $R^2$ & 1.00 / 1.56 / 2.48 & 0.97 / 1.57 / 2.60 \\
$R^2$ pick changed (share of datasets) & --- & 0.19 \\
Fits with log-likelihood higher by $>1$ (share) & --- & 0.37 \\
\addlinespace
\multicolumn{3}{@{}l}{\textit{C. Monte Carlo, S1, adstock-omitted ($R=160$)}} \\
$\Pr(R^2\text{-best} = \text{RoAS-best})$ & 0.294 & 0.287 \\
MAE ratio, $R^2$ pick / RoAS-best & 1.05 & 1.04 \\
Mean RoAS-MAE of the $R^2$ pick & 2.694 & 2.705 \\
Exp-10 / Exp-25 / Exp-50 vs.\ $R^2$ & 0.95 / 0.96 / 0.97 & 0.96 / 0.96 / 0.97 \\
$R^2$ pick changed (share of datasets) & --- & 0.36 \\
Fits with log-likelihood higher by $>1$ (share) & --- & 0.20 \\
\bottomrule
\end{tabular}}
\begin{tablenotes}\footnotesize
\item \textit{Notes}: 20 starts: the same estimator with 20 instead of five random starts from the same start distribution and seeds. Panel A: start at truth higher by $>1$ = the fit started at the generating vector attains a log-likelihood more than one unit above the free fit. Panels B--C: all S1 Monte Carlo datasets refitted with 20 starts; ratios and noisy-benchmark comparisons as in Table~\oldref{tab:mc_cells}, with the same benchmark draws.
\end{tablenotes}
\end{table}
}

{%
\begin{table}[htbp]
\centering
\caption{Monte Carlo Replication of the Main Design: Results by Cell}
\label{tab:mc_cells}
\scriptsize
\resizebox{\textwidth}{!}{%
\begin{tabular}{@{}lllcccccccc@{}}
\toprule
Candidates & Scen. & DGP & $R$ & $\Pr(\text{agree})$ [95\% CI] & $\bar\tau$ & MAE ratio (mean) & MAE ratio (median) & $\Delta R^2$ & Exp-25 / $R^2$ & Median / $R^2$ \\
\midrule
Included & S1 & Hill C & 40 & 0.45 [0.31, 0.60] & 0.76 & 1.40 & 1.06 & 0.000 & 1.01 & 1.71 \\
Included & S1 & Hill S & 40 & 1.00 [0.91, 1.00] & 0.93 & 1.00 & 1.00 & 0.000 & 3.49 & 9.34 \\
Included & S1 & Weibull C & 40 & 0.42 [0.29, 0.58] & 0.73 & 1.36 & 1.15 & 0.000 & 1.03 & 1.57 \\
Included & S1 & Weibull S & 40 & 0.97 [0.87, 1.00] & 0.91 & 1.00 & 1.00 & 0.000 & 2.25 & 3.94 \\
\addlinespace
Included & S2 & Hill C & 40 & 0.50 [0.35, 0.65] & 0.82 & 1.28 & 1.07 & 0.000 & 0.99 & 1.88 \\
Included & S2 & Hill S & 40 & 1.00 [0.91, 1.00] & 0.98 & 1.00 & 1.00 & 0.000 & 3.73 & 11.67 \\
Included & S2 & Weibull C & 40 & 0.70 [0.55, 0.82] & 0.88 & 1.21 & 1.00 & 0.000 & 1.14 & 2.13 \\
Included & S2 & Weibull S & 40 & 1.00 [0.91, 1.00] & 0.92 & 1.00 & 1.00 & 0.000 & 2.34 & 4.65 \\
\addlinespace
Omitted & S1 & Hill C & 40 & 0.30 [0.18, 0.45] & 0.25 & 1.02 & 1.01 & 0.001 & 0.98 & 1.01 \\
Omitted & S1 & Hill S & 40 & 0.33 [0.20, 0.48] & 0.17 & 1.05 & 1.02 & 0.004 & 0.96 & 1.00 \\
Omitted & S1 & Weibull C & 40 & 0.23 [0.12, 0.38] & 0.02 & 1.03 & 1.02 & 0.003 & 0.97 & 1.00 \\
Omitted & S1 & Weibull S & 40 & 0.33 [0.20, 0.48] & 0.11 & 1.06 & 1.04 & 0.007 & 0.95 & 1.02 \\
\addlinespace
Omitted & S2 & Hill C & 40 & 0.35 [0.22, 0.50] & 0.13 & 1.03 & 1.01 & 0.002 & 0.97 & 1.00 \\
Omitted & S2 & Hill S & 40 & 0.38 [0.24, 0.53] & 0.26 & 1.03 & 1.01 & 0.004 & 0.97 & 1.01 \\
Omitted & S2 & Weibull C & 40 & 0.35 [0.22, 0.50] & 0.31 & 1.03 & 1.01 & 0.002 & 0.97 & 1.00 \\
Omitted & S2 & Weibull S & 40 & 0.38 [0.24, 0.53] & 0.31 & 1.03 & 1.01 & 0.008 & 0.97 & 1.05 \\
\addlinespace
\bottomrule
\end{tabular}}
\begin{tablenotes}\scriptsize
\item \textit{Notes}: $R$ = number of independent datasets (each redraws generating parameters, spend, controls and noise). $\Pr(\text{agree})$ = share of datasets in which the highest-$R^2$ candidate is also the lowest-RoAS-MAE candidate (Wilson 95\% interval). $\bar\tau$ = mean Kendall rank correlation between the $R^2$ ranking and the RoAS-accuracy ranking of the four candidates. MAE ratio: mean RoAS-MAE of the $R^2$-selected model divided by the mean RoAS-MAE of the RoAS-best model (Mean), and the median across datasets of the per-dataset ratio (Median). $\Delta R^2$ = mean test $R^2$ of the $R^2$-selected model minus that of the RoAS-best model. Exp-25 / $R^2$ = mean RoAS-MAE obtained when selecting against a noisy benchmark (25\% coefficient of variation per channel) divided by the mean RoAS-MAE of $R^2$ selection; values below one favour the benchmark. Median / $R^2$ = the same ratio of means for the benchmark-free median-consensus diagnostic.
\end{tablenotes}
\end{table}
}

\ifsweeps
{%
\begin{table}[htbp]
\centering
\caption{Robustness to Media-Spend Collinearity: Results by DGP (S2, adstock-included candidates)}
\label{tab:corr_cells}
\scriptsize
\resizebox{\textwidth}{!}{%
\begin{tabular}{@{}llcccccccc@{}}
\toprule
$\rho$ & DGP & $R$ & $\Pr(\text{agree})$ [95\% CI] & $\bar\tau$ & MAE ratio (mean) & MAE ratio (median) & $\Delta R^2$ & Exp-25 / $R^2$ & Median / $R^2$ \\
\midrule
0 & Hill C & 15 & 0.53 [0.30, 0.75] & 0.84 & 1.24 & 1.00 & 0.000 & 1.08 & 1.84 \\
0 & Hill S & 15 & 1.00 [0.80, 1.00] & 0.93 & 1.00 & 1.00 & 0.000 & 3.08 & 5.84 \\
0 & Weibull C & 15 & 0.73 [0.48, 0.89] & 0.91 & 1.09 & 1.00 & 0.000 & 1.22 & 2.05 \\
0 & Weibull S & 15 & 1.00 [0.80, 1.00] & 0.87 & 1.00 & 1.00 & 0.000 & 2.26 & 3.60 \\
0 & \textit{Pooled} & 60 & 0.82 [0.70, 0.89] & 0.89 & 1.07 & 1.00 & 0.000 & 1.76 & 3.00 \\
\addlinespace
0.3 & Hill C & 15 & 0.73 [0.48, 0.89] & 0.91 & 1.22 & 1.00 & 0.000 & 1.08 & 2.82 \\
0.3 & Hill S & 15 & 1.00 [0.80, 1.00] & 1.00 & 1.00 & 1.00 & 0.000 & 2.81 & 7.26 \\
0.3 & Weibull C & 15 & 0.67 [0.42, 0.85] & 0.87 & 1.14 & 1.00 & 0.000 & 1.21 & 1.74 \\
0.3 & Weibull S & 15 & 1.00 [0.80, 1.00] & 0.91 & 1.00 & 1.00 & 0.000 & 2.52 & 3.36 \\
0.3 & \textit{Pooled} & 60 & 0.85 [0.74, 0.92] & 0.92 & 1.09 & 1.00 & 0.000 & 1.78 & 3.26 \\
\addlinespace
0.6 & Hill C & 15 & 0.40 [0.20, 0.64] & 0.76 & 1.29 & 1.06 & 0.000 & 0.94 & 1.79 \\
0.6 & Hill S & 15 & 0.93 [0.70, 0.99] & 0.96 & 1.07 & 1.00 & 0.000 & 2.19 & 4.04 \\
0.6 & Weibull C & 15 & 0.67 [0.42, 0.85] & 0.89 & 1.12 & 1.00 & 0.000 & 1.19 & 1.76 \\
0.6 & Weibull S & 15 & 0.93 [0.70, 0.99] & 0.89 & 1.01 & 1.00 & 0.001 & 2.16 & 3.97 \\
0.6 & \textit{Pooled} & 60 & 0.73 [0.61, 0.83] & 0.87 & 1.10 & 1.00 & 0.000 & 1.62 & 2.84 \\
\addlinespace
0.9 & Hill C & 15 & 0.47 [0.25, 0.70] & 0.71 & 1.35 & 1.05 & 0.000 & 0.88 & 1.03 \\
0.9 & Hill S & 15 & 1.00 [0.80, 1.00] & 0.96 & 1.00 & 1.00 & 0.000 & 1.86 & 4.71 \\
0.9 & Weibull C & 15 & 0.53 [0.30, 0.75] & 0.80 & 1.15 & 1.00 & 0.000 & 1.07 & 1.11 \\
0.9 & Weibull S & 15 & 1.00 [0.80, 1.00] & 0.84 & 1.00 & 1.00 & 0.000 & 1.57 & 2.04 \\
0.9 & \textit{Pooled} & 60 & 0.75 [0.63, 0.84] & 0.83 & 1.08 & 1.00 & 0.000 & 1.38 & 2.12 \\
\addlinespace
\bottomrule
\end{tabular}}
\begin{tablenotes}\scriptsize
\item \textit{Notes}: Columns as in Table~\oldref{tab:mc_cells}. The same random seeds are used at every $\rho$, so differences across rows reflect only the spend correlation; the baseline design uses $\rho=0.3$.
\end{tablenotes}
\end{table}
}

{%
\begin{table}[htbp]
\centering
\caption{Robustness to Noise Level: Results by DGP (S2, adstock-included candidates)}
\label{tab:noise_cells}
\scriptsize
\resizebox{\textwidth}{!}{%
\begin{tabular}{@{}llcccccccc@{}}
\toprule
$\sigma$ & DGP & $R$ & $\Pr(\text{agree})$ [95\% CI] & $\bar\tau$ & MAE ratio (mean) & MAE ratio (median) & $\Delta R^2$ & Exp-25 / $R^2$ & Median / $R^2$ \\
\midrule
0.02 & Hill C & 15 & 0.67 [0.42, 0.85] & 0.89 & 1.18 & 1.00 & 0.000 & 1.09 & 2.07 \\
0.02 & Hill S & 15 & 1.00 [0.80, 1.00] & 1.00 & 1.00 & 1.00 & 0.000 & 5.47 & 11.78 \\
0.02 & Weibull C & 15 & 0.47 [0.25, 0.70] & 0.82 & 1.33 & 1.05 & 0.000 & 1.09 & 1.81 \\
0.02 & Weibull S & 15 & 1.00 [0.80, 1.00] & 0.89 & 1.00 & 1.00 & 0.000 & 2.86 & 8.71 \\
0.02 & \textit{Pooled} & 60 & 0.78 [0.66, 0.87] & 0.90 & 1.17 & 1.00 & 0.000 & 1.86 & 4.37 \\
\addlinespace
0.05 & Hill C & 15 & 0.60 [0.36, 0.80] & 0.73 & 1.34 & 1.00 & 0.000 & 0.89 & 1.36 \\
0.05 & Hill S & 15 & 1.00 [0.80, 1.00] & 1.00 & 1.00 & 1.00 & 0.000 & 2.64 & 5.74 \\
0.05 & Weibull C & 15 & 0.33 [0.15, 0.58] & 0.69 & 1.25 & 1.15 & 0.000 & 1.03 & 1.09 \\
0.05 & Weibull S & 15 & 1.00 [0.80, 1.00] & 0.89 & 1.00 & 1.00 & 0.000 & 1.75 & 3.98 \\
0.05 & \textit{Pooled} & 60 & 0.73 [0.61, 0.83] & 0.83 & 1.15 & 1.00 & 0.000 & 1.38 & 2.51 \\
\addlinespace
0.1 & Hill C & 15 & 0.33 [0.15, 0.58] & 0.44 & 1.44 & 1.06 & 0.001 & 0.78 & 1.25 \\
0.1 & Hill S & 15 & 0.93 [0.70, 0.99] & 0.98 & 1.01 & 1.00 & 0.000 & 1.83 & 4.09 \\
0.1 & Weibull C & 15 & 0.60 [0.36, 0.80] & 0.56 & 1.43 & 1.00 & 0.000 & 0.84 & 1.05 \\
0.1 & Weibull S & 15 & 0.87 [0.62, 0.96] & 0.84 & 1.07 & 1.00 & 0.001 & 1.28 & 2.18 \\
0.1 & \textit{Pooled} & 60 & 0.68 [0.56, 0.79] & 0.71 & 1.22 & 1.00 & 0.000 & 1.11 & 1.86 \\
\addlinespace
0.2 & Hill C & 15 & 0.47 [0.25, 0.70] & 0.42 & 1.27 & 1.03 & 0.002 & 0.84 & 1.09 \\
0.2 & Hill S & 15 & 0.60 [0.36, 0.80] & 0.82 & 1.10 & 1.00 & 0.001 & 1.33 & 1.78 \\
0.2 & Weibull C & 15 & 0.47 [0.25, 0.70] & 0.07 & 1.53 & 1.19 & 0.002 & 0.72 & 0.92 \\
0.2 & Weibull S & 15 & 0.53 [0.30, 0.75] & 0.78 & 1.23 & 1.00 & 0.002 & 0.99 & 2.00 \\
0.2 & \textit{Pooled} & 60 & 0.52 [0.39, 0.64] & 0.52 & 1.30 & 1.00 & 0.002 & 0.93 & 1.51 \\
\addlinespace
\bottomrule
\end{tabular}}
\begin{tablenotes}\scriptsize
\item \textit{Notes}: Columns as in Table~\oldref{tab:mc_cells}. The same random seeds are used at every $\sigma$, so differences across rows reflect only the noise level; the baseline design uses $\sigma=0.02$.
\end{tablenotes}
\end{table}
}

{%
\begin{table}[htbp]
\centering
\caption{Adstock-Form Misspecification: Results by DGP (S2; delayed-adstock DGP, geometric-adstock candidates)}
\label{tab:adstock_cells}
\scriptsize
\resizebox{\textwidth}{!}{%
\begin{tabular}{@{}lcccccccc@{}}
\toprule
DGP & $R$ & $\Pr(\text{agree})$ [95\% CI] & $\bar\tau$ & MAE ratio (mean) & MAE ratio (median) & $\Delta R^2$ & Exp-25 / $R^2$ & Median / $R^2$ \\
\midrule
Hill C & 15 & 0.40 [0.20, 0.64] & 0.44 & 1.24 & 1.08 & 0.001 & 0.89 & 1.30 \\
Hill S & 15 & 0.40 [0.20, 0.64] & 0.27 & 1.35 & 1.12 & 0.008 & 0.83 & 1.08 \\
Weibull C & 15 & 0.33 [0.15, 0.58] & 0.38 & 1.58 & 1.32 & 0.002 & 0.70 & 0.84 \\
Weibull S & 15 & 0.27 [0.11, 0.52] & 0.20 & 1.21 & 1.10 & 0.016 & 0.90 & 1.00 \\
\addlinespace
\textit{Pooled} & 60 & 0.35 [0.24, 0.48] & 0.32 & 1.30 & 1.11 & 0.007 & 0.85 & 1.01 \\
\bottomrule
\end{tabular}}
\begin{tablenotes}\scriptsize
\item \textit{Notes}: Columns as in Table~\oldref{tab:mc_cells}. Sales are generated with a delayed adstock kernel $w_k \propto \alpha^{(k-\theta)^2}$ ($k=0,\dots,12$), peak lag $\theta\in\{1,2\}$ drawn per channel and total carryover matched to the geometric kernel; all candidates use geometric adstock, so carryover is misspecified for every candidate while the true saturation family remains among them.
\end{tablenotes}
\end{table}
}

\begin{figure}[htbp]
    \centering
    \includegraphics[width=\textwidth]{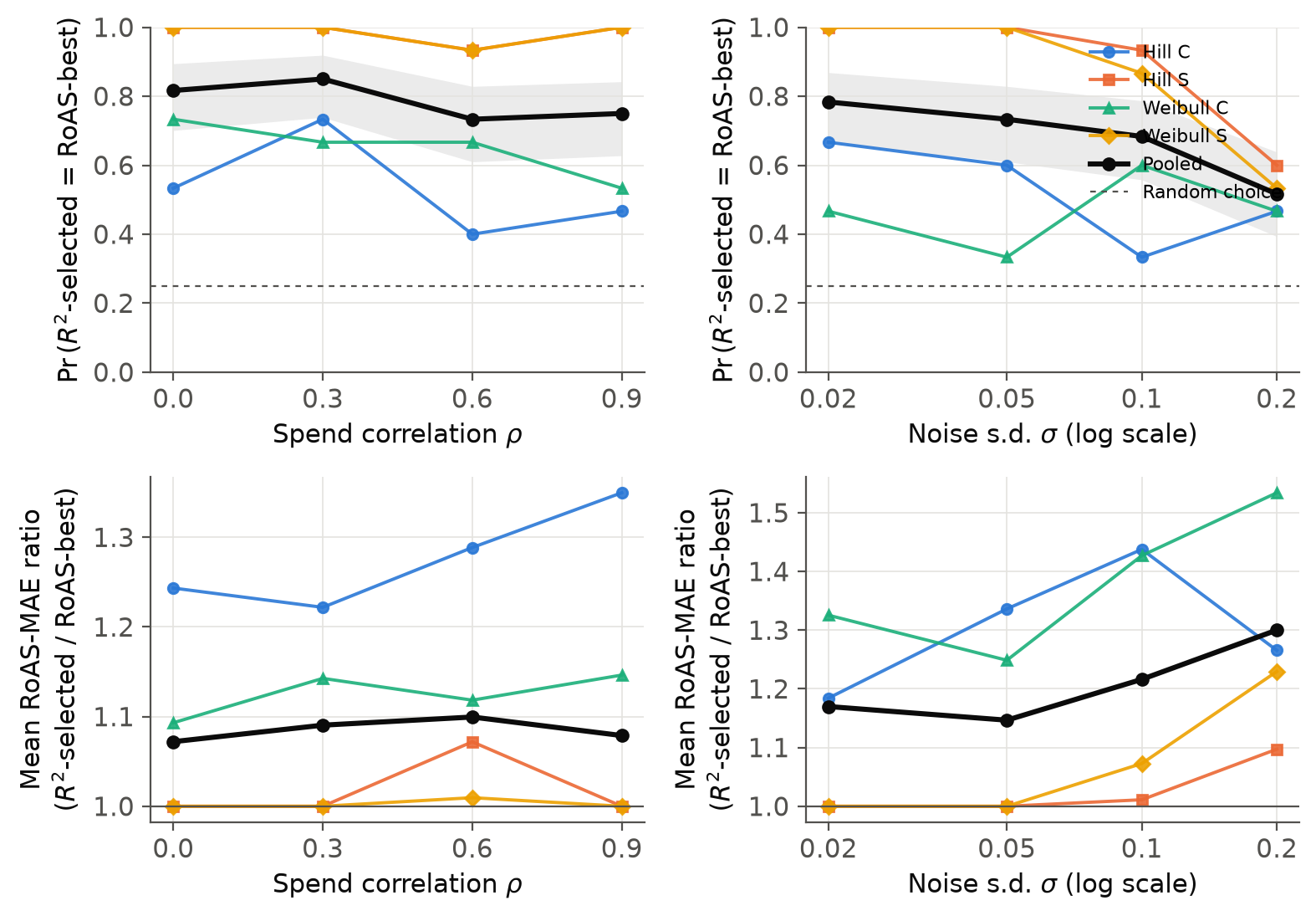}
    \caption{Collinearity (left) and noise (right) studies, S2 with adstock-included candidates. Top: share of datasets in which the highest-$R^2$ candidate is also the RoAS-best candidate (pooled line with 95\% Wilson band; dashed line = random choice among four candidates). Bottom: mean RoAS-MAE of the $R^2$-selected model relative to the RoAS-best model. Colored lines show individual DGPs; black lines pool the four DGPs.}
    \label{fig:sweeps}
\end{figure}
\fi

{%
\begin{table}[htbp]
\centering
\caption{Noisy-Benchmark Selection and Consensus Diagnostics: Paired Comparisons with $R^2$ Selection}
\label{tab:selection_detail}
\scriptsize
\resizebox{\textwidth}{!}{%
\begin{tabular}{@{}lcccccccc@{}}
\toprule
& \multicolumn{3}{c}{\textbf{Exp-25 vs.\ $R^2$ selection}} & \multicolumn{2}{c}{\textbf{Exp-50 vs.\ $R^2$}} & \multicolumn{3}{c}{\textbf{Median consensus / $R^2$}} \\
\cmidrule(lr){2-4}\cmidrule(lr){5-6}\cmidrule(lr){7-9}
Study & MAE ratio [95\% CI] & $\Pr$(better) & $\Pr$(worse by $>$10\%) & MAE ratio [95\% CI] & $\Pr$(worse by $>$10\%) & Midpoint & Upper & Leave-one-out change \\
\midrule
Monte Carlo, S1, adstock-included & 1.56 [1.40, 1.76] & 0.16 & 0.32 & 2.48 [2.16, 2.89] & 0.46 & 2.93 & 4.73 & 0.34 \\
Monte Carlo, S2, adstock-included & 1.74 [1.57, 1.93] & 0.13 & 0.31 & 3.07 [2.69, 3.52] & 0.46 & 3.80 & 7.47 & 0.34 \\
Monte Carlo, S1, adstock-omitted & 0.96 [0.95, 0.97] & 0.68 & 0.00 & 0.97 [0.95, 0.98] & 0.02 & 1.01 & 0.98 & 0.36 \\
Monte Carlo, S2, adstock-omitted & 0.97 [0.97, 0.98] & 0.63 & 0.00 & 0.98 [0.97, 0.99] & 0.01 & 1.03 & 0.99 & 0.35 \\
Collinearity, $\rho=0$ & 1.76 [1.56, 2.01] & 0.11 & 0.32 & 3.00 [2.50, 3.65] & 0.48 & 3.00 & 7.94 & 0.35 \\
Collinearity, $\rho=0.3$ & 1.78 [1.55, 2.08] & 0.09 & 0.34 & 3.21 [2.62, 3.97] & 0.49 & 3.26 & 7.79 & 0.36 \\
Collinearity, $\rho=0.6$ & 1.62 [1.46, 1.83] & 0.17 & 0.30 & 2.77 [2.33, 3.35] & 0.45 & 2.84 & 6.56 & 0.40 \\
Collinearity, $\rho=0.9$ & 1.38 [1.27, 1.52] & 0.16 & 0.22 & 2.42 [2.12, 2.84] & 0.39 & 2.12 & 3.96 & 0.36 \\
Noise, $\sigma=0.02$ & 1.86 [1.56, 2.24] & 0.13 & 0.30 & 3.47 [2.74, 4.39] & 0.46 & 4.37 & 7.64 & 0.33 \\
Noise, $\sigma=0.05$ & 1.38 [1.22, 1.59] & 0.17 & 0.30 & 2.28 [1.91, 2.76] & 0.45 & 2.51 & 4.73 & 0.36 \\
Noise, $\sigma=0.1$ & 1.11 [0.99, 1.27] & 0.24 & 0.28 & 1.63 [1.39, 1.94] & 0.42 & 1.86 & 2.97 & 0.35 \\
Noise, $\sigma=0.2$ & 0.93 [0.81, 1.06] & 0.36 & 0.18 & 1.18 [1.04, 1.34] & 0.30 & 1.51 & 2.04 & 0.36 \\
Adstock form, delayed-adstock DGP & 0.85 [0.77, 0.92] & 0.50 & 0.12 & 0.93 [0.85, 1.02] & 0.21 & 1.01 & 1.08 & 0.38 \\
\bottomrule
\end{tabular}}
\begin{tablenotes}\scriptsize
\item \textit{Notes}: All rules select among the same four fitted candidates of each dataset and are scored against the known RoAS. MAE ratio: mean RoAS-MAE of the rule divided by the mean RoAS-MAE of the $R^2$-selected model, with a 95\% bootstrap interval over datasets (benchmark-noise draws are averaged within each dataset first). $\Pr$(better) and $\Pr$(worse by $>$10\%): share of dataset--draw pairs in which the benchmark-selected model has lower RoAS-MAE than, or RoAS-MAE more than 10\% above, the $R^2$-selected model. Median consensus: ratio of mean RoAS-MAE for the candidate closest to the componentwise median RoAS vector, with the median of four values taken as the mean of the two central values (midpoint) or the upper central value (upper). Leave-one-out change: share of removals of one non-selected candidate that change the median-consensus selection.
\end{tablenotes}
\end{table}
}

\end{document}